\documentclass[twocolumn]{aastex631}

\usepackage{color}
\usepackage{url}
\usepackage{hyperref}
\usepackage{xspace}
\usepackage{soul}
\usepackage{multirow,threeparttable}
\usepackage{amsmath} 
 
\newcommand{\psrtar}{SRGA\,J144459.2$-$604207}

\newcommand{\nustar}{{NuSTAR}\xspace}
\newcommand{\ep}{{EP}\xspace}
\newcommand{\epfull}{{Einstein Probe}\xspace}
\newcommand{\nicer}{{NICER}\xspace}  
\newcommand{\cxo}{{\it{Chandra}}\xspace}
\newcommand{\xmm}{{XMM-Newton}\xspace}
\newcommand{\swift}{{Swift}\xspace}
\newcommand{\Integ}{{INTEGRAL}\xspace}  
\newcommand{\maxi}{{MAXI}\xspace}  
\newcommand{\rxte}{{\it{RXTE}}\xspace}
\newcommand{\hxmt}{{{\it Insight}-HXMT}\xspace}
\newcommand{\ixpe}{{IXPE}\xspace}

\def\chiq{$\chi^2$}

\def\be{\begin{equation}} 
\def\ee{\end{equation}}

\begin{document}

\title{Timing and spectral studies of \psrtar\ with \nicer, \epfull, \ixpe, \nustar, \hxmt\ and \Integ  during its 2024 outburst}

\correspondingauthor{Zhaosheng Li}
\email{lizhaosheng@xtu.edu.cn}

\correspondingauthor{Yong Chen}
\email{ychen@ihep.ac.cn}

\author[0000-0003-2310-8105]{Zhaosheng Li}
\affiliation{Key Laboratory of Stars and Interstellar Medium, Xiangtan University, Xiangtan 411105, Hunan, China}

\author[0000-0002-7889-6586]{Lucien Kuiper}
\affiliation{SRON - Space Research Organisation Netherlands, Niels Bohrweg 4, 2333 CA, Leiden, The Netherlands}

\author{Yuanyue Pan}
\affiliation{Key Laboratory of Stars and Interstellar Medium, Xiangtan University, Xiangtan 411105, Hunan, China}

\author[0000-0002-9042-3044]{Renxin Xu}
\affiliation{Department of Astronomy, School of Physics, Peking University, Beijing 100871, China}
\affiliation{Kavli Institute for Astronomy and Astrophysics, Peking University, Beijing 100871, China}

\author[0000-0001-9834-2196]{Yong Chen}
\affiliation{Key Laboratory of Particle Astrophysics, Institute of High Energy Physics, Chinese Academy of Sciences, 19B Yuquan Road, Beijing 100049, China}

\author[0000-0002-3776-4536]{Mingyu Ge}
\affiliation{Key Laboratory of Particle Astrophysics, Institute of High Energy Physics, Chinese Academy of Sciences, 19B Yuquan Road, Beijing 100049, China}

\author{Yue Huang}
\affiliation{Key Laboratory of Particle Astrophysics, Institute of High Energy Physics, Chinese Academy of Sciences, 19B Yuquan Road, Beijing 100049, China}

\author{Shumei Jia}
\affiliation{Key Laboratory of Particle Astrophysics, Institute of High Energy Physics, Chinese Academy of Sciences, 19B Yuquan Road, Beijing 100049, China}

\author{Xiaobo Li}
\affiliation{Key Laboratory of Particle Astrophysics, Institute of High Energy Physics, Chinese Academy of Sciences, 19B Yuquan Road, Beijing 100049, China}

\author[0000-0003-0274-3396]{Liming Song}
\affiliation{Key Laboratory of Particle Astrophysics, Institute of High Energy Physics, Chinese Academy of Sciences, 19B Yuquan Road, Beijing 100049, China}

\author{Jinlu Qu}
\affiliation{Key Laboratory of Particle Astrophysics, Institute of High Energy Physics, Chinese Academy of Sciences, 19B Yuquan Road, Beijing 100049, China}

\author{Shu Zhang}
\affiliation{Key Laboratory of Particle Astrophysics, Institute of High Energy Physics, Chinese Academy of Sciences, 19B Yuquan Road, Beijing 100049, China}

\author{Lian Tao}
\affiliation{Key Laboratory of Particle Astrophysics, Institute of High Energy Physics, Chinese Academy of Sciences, 19B Yuquan Road, Beijing 100049, China}

\author{Hua Feng}
\affiliation{Key Laboratory of Particle Astrophysics, Institute of High Energy Physics, Chinese Academy of Sciences, 19B Yuquan Road, Beijing 100049, China}

\author[0000-0001-5586-1017]{Shuang-Nan Zhang}
\affiliation{Key Laboratory of Particle Astrophysics, Institute of High Energy Physics, Chinese Academy of Sciences, 19B Yuquan Road, Beijing 100049, China}

\author[0000-0003-3095-6065]{Maurizio Falanga}
\affiliation{International Space Science Institute (ISSI), Hallerstrasse 6, 3012 Bern, Switzerland}
\affiliation{Physikalisches Institut, University of Bern, Sidlerstrasse 5, 3012 Bern, Switzerland}

\begin{abstract}
\psrtar\ is a newly confirmed accreting millisecond X-ray pulsar and type I X-ray burster. We present the broadband X-ray timing and spectral behaviors of \psrtar\ during its 2024 outburst. The data were collected from \nicer, \epfull, \ixpe, \hxmt, \nustar and \Integ observations. X-ray pulsations have been detected for the 1.5--90 keV energy range throughout the `ON' phase of the outburst from MJD $\sim 60355-60385$. We refined the orbital and spin ephemerides assuming a circular orbit, and found that the pulsar was in a spin-up state during MJD $\sim$ 60361--60377 showing a significant spin-up rate $\dot{\nu}$ of $(3.15\pm 0.36)\times10^{-13}~{\rm Hz~s^{-1}}$. Around MJD $\sim 60377$ a swing was detected in the spin evolution accompanied by significantly enhanced pulsed emission. We studied the pulse profile morphology during the X-ray bursts as observed by \hxmt, \ixpe and \nustar. During the bursts, pulsations were detected across the 2--60 keV with shapes broadly consistent with those observed for the persistent emission. We found, however, that the `burst' pulse profiles exhibit significant phase offsets relative to the pre- and post-burst profiles. These offsets systematically decrease with increasing energy, $\Delta \phi\approx0.15$, 0.11 and 0.02 for \ixpe, \hxmt ME and HE in 2--8, 5--30 and 20--60 keV, respectively, and $\Delta \phi\approx0.21$, 0.10 and 0.07 for \nustar in 3--10, 20--35 and 35--60 keV, respectively, compared to the pre- and post-burst profiles.  We performed a joint spectral analysis of quasi-simultaneous \nicer, \nustar, and \hxmt data for two epochs. The resulting spectra from both observations were consistent and well-described by an absorbed thermal Comptonization model, \texttt{nthcomp},  plus relativistic reflection, \texttt{relxillCp}.

\end{abstract}


\section{Introduction} \label{sec:intro}

Accreting millisecond X-ray pulsars (AMXPs) host a fast-rotating neutron star (NS) and a low-mass companion in the main sequence, belonging to a sub-class of NS low-mass X-ray binary \citep[LMXB; see e.g.,][for reviews]{Patruno2012,Salvo22}. The strong magnetic field of NS, i.e., a typical strength of $10^8-10^9$ G, in an AMXP, channels the inflowing matter from the inner accretion disk onto the NS surface at the magnetic poles. This accretion process produces hot spots on the NS surface, which in turn generate soft X-ray pulsations as the NS rotates. Meanwhile, the soft X-ray photons from hot spots are up-scattered by the in-falling accretion column and emit hard X-ray pulsation above 100 keV, as observed by \rxte, \Integ, and \hxmt\ in the last two decades \citep{falanga05,falanga05b,falanga08,falanga11,falanga12,Falanga07,deFalcoa,deFalcob,Kuiper20,ZLi21,ZLi23,ZLi24}.   As the accreted material accumulates at the magnetic poles, it eventually spreads over the entire NS surface. This spreading layer can be triggered via unstable thermonuclear burning of accreted helium or a mixture of hydrogen and helium, which is also known as type I X-ray burst characterized by the rapid releasing energy of $10^{39}-10^{40}$ ergs in 10--100 s \citep[see e.g.,][for reviews]{galloway08,Galloway21}.

\psrtar\ was discovered on February 21, 2024 as a new bright Galactic transient by SRG ART-XC during scans of an ongoing all-sky survey \citep{2024ATel16464,Molkov24}. In subsequent \nicer\ observations coherent X-ray pulsations were discovered at $\sim 447.9$ Hz, confirming the source as an AMXP \citep{2024ATel16480,Ng24, Papitto25}. Follow-up optical and radio observations were carried out. The radio emission from \psrtar\ has been detected by ATCA by using its \cxo\ position \citep{2024ATel16510}, resulting in the best-determined source location of $\alpha_{\rm 2000} = 14^{\rm h}44^{\rm m}59\fs0(2)$ and $\delta_{\rm 2000} = -60\degr41\arcmin56\farcs1(4)$ \citep{2024ATel16511}. None of the optical or near-infrared counterparts have been found at the ATCA location \citep{2024ATel16476,2024ATel16477,2024ATel16487}. Polarized emission from \psrtar, with an average polarization degree of $2.3\% \pm 0.4\%$  at an angle of $59^\circ\pm6^\circ$, was recently reported based on observations by the Imaging X-ray Polarimetry Explorer \citep[IXPE;][]{Papitto25}. Joint XMM-Newton and NuSTAR broadband spectral analysis of \psrtar\ reveals prominent relativistically blurred reflection features, including a broadened iron emission line and a blueshifted Fe XXVI absorption edge \citep{Malacaria25}.

Several X-ray telescopes, that is, \hxmt, \swift, \cxo, \nustar, \Integ, NinjaSAT and \ixpe, have detected quasi-periodic thermonuclear X-ray bursts from \psrtar\ with burst recurrence times increasing from 1.5 to 10.0 h as as the persistent emission decreases \citep{2024ATel16548,2024ATel16475,2024ATel16510,2024ATel16476,2024ATel16495,2024ATel16485,Papitto25,Fu25, Malacaria25}. \citet{Fu25} found an anti-correlation between the recurrence time and the local mass accretion rate, $\Delta T_{\rm rec}\sim\dot{m}^{-0.91\pm0.02}$. A similar relation has also been reported for AMXPs Swift J1748.9--2021 and MAXI J1816--195 \citep{ZLI2018,Chen22}. The distance to \psrtar\ is estimated at about 10 kpc \citep{Fu25}, a value derived from Photospheric Radius Expansion (PRE) bursts observed by \hxmt. In these bursts, intense luminosity causes the NS's photosphere to temporarily expand due to radiation pressure, with the burst emission reaching the local Eddington limit. By assuming the observed peak flux during this phase corresponds to the Eddington luminosity, which acts as a standard candle \citep{kuulkers03}, the distance to \psrtar\ was determined \citep{Fu25}.

In this work, we analyze the 2024 outburst data of \psrtar\ collected by \hxmt, \nicer, \nustar, \Integ, \ixpe and \epfull, as described in Sect.~\ref{sec:data}. We present the broadband timing and spectral characteristics of \psrtar\ in Sects.~\ref{sec:timing} and \ref{sec:spectra}, and finally the results are discussed in Sect.~\ref{sec:diss}. 

\section{Data Reduction} \label{sec:data}

\subsection{\hxmt\ observations}\label{sec:hxmt}
\hxmt\ \citep[Insight Hard X-ray Modulation Telescope,][]{hxmt} is the first Chinese X-ray telescope, and is equipped with three slat-collimated instruments: the Low Energy X-ray telescope \citep[LE, 1--12 keV; ][]{hxmt-le}, the Medium Energy X-ray telescope \citep[ME, 5--35 keV; ][]{hxmt-me} and the High Energy X-ray telescope \citep[HE, 20--350 keV; ][]{hxmt-he}, providing capabilities for broadband X-ray timing and spectroscopy \citep{ZLi21,ZLi23,ZLi24}. 

\hxmt\ carried out high-cadence observations of \psrtar\ starting on MJD 60363.294,  around the outburst peak revealed by MAXI. The set of 53 observations includes runs P0614373001 -- P0614373006.
The LE, ME and HE data were used to investigate the broadband spectral properties. However, due to the limited time resolution, $\sim 1$ ms,  of the LE \citep{Tuo22}, only ME and HE data were used to perform the timing analysis.

We analyzed  the data using the \hxmt\
Data Analysis Software  (HXMTDAS) version 2.05. The LE, ME and HE data were calibrated by using the scripts {\tt lepical},  {\tt mepical} and {\tt hepical}, respectively.  The good time intervals were individually selected from the scripts {\tt legtigen}, {\tt megtigen} and {\tt hegtigen} for LE, ME, and HE, respectively, with the standard criteria, including the earth elevation angle, ELV $> 10^\circ$, the cutoff rigidity, COR $> 8$ GeV, the satellite located outside the South Atlantic Anomaly region longer than 300 s, and the offset angle from the pointing direction smaller than $0\farcs04$. 

From the cleaned 1 s binned light curves, type I X-ray bursts were identified and removed in the timing and spectral analysis. No bursts were shown in HE data because the burst emissions are mainly dominated by soft X-ray photons. Background subtracted light curves for the LE, ME and HE were generated (see Fig.~\ref{fig:outburst}).

The spectra and their response matrix files are produced by the tools {\tt hespecgen} and {\tt herspgen} for HE, {\tt mespecgen} and {\tt merspgen} for ME, and {\tt lespecgen} and {\tt lerspgen} for LE, respectively. Finally, we obtained the cleaned events using {\tt mescreen} and {\tt hescreen} and  barycentered with the tool {\tt hxbary}.

\subsection{\nicer\ observations}\label{sec:nicer}
From the public HEASARC archive, we found that \nicer\ \citep{nicer} observed \psrtar\ between February 21, 2024 19:56:30 and May 3, 2024 20:49:40 (MJD 60361.83--60433.87).\footnote{All observation start and stop times reported in this work are given in Coordinated Universal Time (UTC) unless explicitly stated otherwise.} The source went off around March 14, 2024 21:18 (MJD 60383.89) after which the pulsations became undetectable and the source entered the off state. 

The total exposure time during its active period amounts 62.9 ks  using the calibrated unfiltered (UFA) event files. We followed the standard data analysis to extract the cleaned event files using \texttt{nicer-l2}. The cleaned light curves were extracted using the tool \texttt{nicerl3-lc}. From the light curves, five type I X-ray bursts were detected, including the one reported in \citet{Ng24}. After removing the time intervals that cover these bursts, we generated the source and 3C50 background spectra, the arf, and the response files from the command \texttt{nicerl3-spect}. We verified our spectral results by re-extracting the background with the SCORPEON model. This yielded spectral parameters consistent with those from the 3C50 model, indicating that our results are not sensitive to the choice of background model. The spectra were optimally grouped by the tool {\tt ftgrouppha}. Due to light leakage issues of \nicer, the exposure time of cleaned event files was reduced to only 38.7 ks. To better cover the outburst, we extracted the 0.5--10 and 12--15 keV light curves directly from the UFA event files.  From the 12--15 keV light curve,  we identified time intervals containing flaring particle background and ignored these constructing the 0.5--10 keV light curve (see the third panel in Fig.~\ref{fig:outburst}). 

\subsection{\ixpe\ observations}\label{sec:ixpe}

\begin{figure} 
\centering
\includegraphics[width=\linewidth]{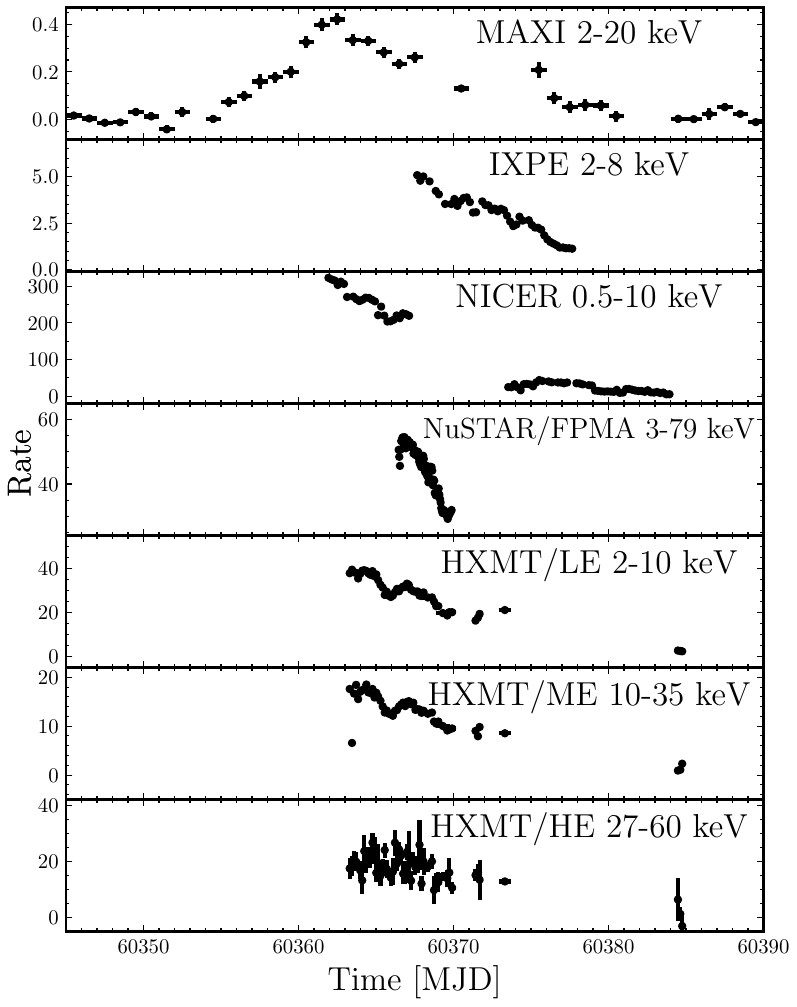}
\caption{The light curves of \psrtar\ during its 2024 outburst. From top to bottom: the background subtracted light curves from \maxi\ (1 day binned) in units of $\rm{photons~cm^{-2}~s^{-1}}$, \ixpe\ (0.2 day binned), \nicer\ (0.2 day binned),  \nustar/FPMA (1 hr binned), and \hxmt\ LE/ME/HE (each point represents an exposure observation)  are displayed, respectively. The energy range of each light curve is indicated in each panel. Time intervals of \nicer, \ixpe, \hxmt\ LE and ME, and \nustar\ covering X-ray bursts are removed. The \nicer\ light curves are from the UFA event files in 0.5--10 keV.
}
\label{fig:outburst}
\end{figure}

\ixpe\ \citep{ixpe} is an X-ray telescope equipped with three identical detector units (DUs) providing imaging, polarization and spectral capabilities, while maintaining a high-time resolution of better than 100 $\mu {\rm s}$.  \ixpe\ carried out a ToO observation \citep[PI: A. Papitto;][]{Papitto25} of \psrtar\ between February 27, 2024 13:09:48 and March 8, 2024 16:24:11 (MJD 60367.55 -- 60377.68) for a net exposure time of $\sim553$ ks. We combined the data collected by all three DUs from \ixpe\ Level-2 files. We extracted the events from a circular region centered at the source position with a radius of $100\arcsec$, and the background events from a region centered at $(\alpha_{2000},\delta_{2000})= (14^{\rm h}45^{\rm m}17\fs46, -60\degr37\arcmin33\farcs92)$.  Thanks to the large duty cycle of about 63\%, \ixpe\ detected 52 X-ray bursts, identified from the 1 s binned light curve, with a recurrence time increasing from 2.2 hr to 7.9 hr as the persistent count rate decreased during the fading part of the outburst \citep[see also][]{Papitto25}. The outburst light curve with X-ray bursts removed is shown in the second panel in Fig.~\ref{fig:outburst}. 

\subsection{\nustar\ observations}\label{sec:nustar}
 
On February 26, 2024 11:01:06 \nustar\ \citep{nustar} started a ToO observation of \psrtar\ for a total exposure time of  157.7 ks (Obs. ID 80901307002; MJD 60366.46--60369.88). The event files from the FPMA and FPMB focal plane modules have been cleaned using the \nustar\ pipeline tool \texttt{nupipeline}. The source light curves were extracted from a circle region with a radius of $200\arcsec$ centered on the source location using \texttt{nuproducts}. From the light curve, 23 type I X-ray bursts could be identified during the \nustar\ observation with the recurrence times ranging from 1.97 to 2.85 hr \citep[see also,][]{Papitto25, Malacaria25}. Moreover, we also identified particle flares, which showed sharp peaks in the 3--79 keV light curves with count rates exceeding 100~cnt~s$^{-1}$. After removing the bursts and flares, the persistent count rate in the 3--79 keV band of \nustar\ decreased from $\sim$52 to 44~cnt~s$^{-1}$ in $10^4$ s, increased to a peak of 56 ~cnt~s$^{-1}$ in next $10^4$ s, and then followed a slowly decreasing trend to 30 ~cnt~s$^{-1}$ superposed with some fluctuation (see the middle panel of Fig.~\ref{fig:outburst}).

To perform joint spectral fitting with \nicer\ spectra (see Sect.~\ref{sec:spectra}), we excluded the time intervals of the 23 X-ray bursts and other flares in producing the source spectra, response, and ancillary response files. The background spectra were obtained from a source free circular region with a radius of $100\arcsec$ centered on $(\alpha_{2000},\delta_{2000})= (14^{\rm h}45^{\rm m}55\fs35, -60\degr36\arcmin01\farcs59)$. 

\subsection{\Integ\ observations}

On February 24, 2024 09:31:21 \Integ\ \citep{winkler03} started ToO observations of \psrtar\ during orbital revolution 2747 for about 60 ks (PI: E. Kuulkers). Also, during the next consecutive revolutions, 2748 and 2749, the source was observed for 100.8 and 90 ks, respectively. The initial ToO observations performed during the active period of the source cover the time range MJD 60364.396 -- 60370.466 (Feb. 24, 2024 09:31:21 -- Mar. 1, 2024 11:25:14), nicely overlapping with concurrent \nicer, \hxmt, \nustar\ and \ixpe\ observations.
Later, during the off-state of the source \Integ\ performed two more dedicated ToO observations of \psrtar, starting at Mar. 14, 2024 11:54:10 (MJD 60383.496), during revolutions 2754 and 2756 for 90 and 89 ks, respectively. Moreover, the source was in the fully-coded soft gamma-ray imager ISGRI \citep{ubertini03} field-of-view during ToO observations of an unrelated Galactic transient, Swift J151857.0-5721, for revolutions 2755, 2757--2761, all performed when \psrtar\ had already entered the `OFF' state. 

In this work, driven by sensitivity considerations, we only performed a timing analysis using data from the soft gamma-ray coded mask imager ISGRI (20-300 keV) aboard \Integ\ collected during revolutions 2747--2749 at the time of its active `ON' period.

For the ISGRI timing analysis we used only observations for which the source off-axis angle was less than $14\fdg5$. To remove flaring events a filter was applied to the \Integ\ ISGRI count rate distribution, excluding excursions in excess of $4\sigma$ above the median value.  
Moreover, the events coming from non-noisy detector pixels had to satisfy some criteria: 1) the event rise time should be within 7--90, and 2) the pixel illumination factor (PIF) must be in the range 0.25--1 (i.e. more than 25\% of a detector pixel is illuminated by the source) to reduce the background. 

The main goal was to investigate until what energy the pulsations could be detected and to study possible pulse-shape morphology changes as a function of energy.

\subsection{Einstein Probe observations}\label{sec:ep}

The Einstein Probe \citep[\ep;][]{EP2025}  was launched on 9 January 2024 with on board the Wide-field X-ray Telescope (WXT, 0.5--4 keV) and the Follow-up X-ray Telescope \citep[FXT, 0.3--10 keV;][]{EP-FXT}. The FXT consists of two pn-CCD modules, FXT-A and FXT-B, which can operate in Full-Frame Mode (FF), the Partial-Window Mode (PW) and the Timing Mode (TM). In TM the time resolution is about $46~\mu s$ \citep{EP-FXT-Timing}.  During the on-orbit calibration phase, \ep\ observed \psrtar\ twice (PI: Y. Chen), namely at observation Obs. 136000051117 (MJD 60382.36--60383.06) and Obs. 13600005118 (MJD 60383.36--60384.33), both near the end of the `ON' state of the source. The first observation was carried out in PW and TM, while the second observation was conducted solely in FF. To search for pulsations we therefore only focus on the first observation, Obs. 136000051117, in TM. The data were processed following the standard procedures embedded in the Follow-up X-ray Telescope Data Analysis Software (FXTDAS) version 1.10 using the tool \texttt{fxtchain}. The exposure times for both FXT-A and FXT-B were about 7.365 ks, and no X-ray bursts were observed. The TM data were barycentered using the \texttt{fxtbary} %
procedure adopting the DE405 solar system ephemeris and the ATCA radio-position of \psrtar.

\begin{table}[t] 
{\small
\caption{The orbital and spin parameters of \psrtar\ as derived in this work using \nicer, \hxmt\, ME and \ixpe\ data along with the ATCA radio location \citep{2024ATel16511}.}
\centering
\begin{tabular}{lcc} 
\hline \hline 
Parameter                      & Values                           &Units    \\
\hline 
\noalign{\smallskip}  
$\alpha_{2000}$                & $14^{\hbox{\scriptsize h}} 44^{\hbox{\scriptsize m}} 59\fs0$ &         \\   
$\delta_{2000}$                & $-60\degr41\arcmin56\farcs1$     &         \\         
\noalign{\smallskip}  
\hline
\multicolumn{3}{c}{Constant Frequency \nicer\ model (4d-{\tt SIMPLEX})}\\
\hline  
\noalign{\smallskip}
JPL Ephemeris                  & DE405                            &         \\    
Validity range                 & 60361.83-- 60383.89              &MJD (TDB)\\ 
$ e $                          & $0$ (fixed)                      &         \\
$ P_{\rm orb} $                & 18803.670\,6(30)                 &s        \\           
$ a_{\rm x}\, \sin i$          & 0.650\,513(20)                   &lt-s     \\      
$ T_{\rm asc} $                & 60361.858\,933\,2(15)            &MJD (TDB)\\
$t_0$ (Epoch)                  & 60373                            &MJD (TDB)\\
$\nu$                          & 447.871\,561\,224(11)            &Hz       \\
\noalign{\smallskip}
\hline
\multicolumn{3}{c}{Constant Frequency model (ToA)}\\
\hline 
Validity range                 & 60361-- 60377                 &MJD (TDB)\\                      
$t_0$ (Epoch)                  & 60364                         &MJD (TDB)\\
$\nu$                          & 447.871\,561\,272\,4(48)      &Hz       \\

\chiq{}/d.o.f                  & 150.65/(67-1)=2.28            &         \\
\noalign{\smallskip}
\hline
\multicolumn{3}{c}{Spin-up model (ToA)}\\
\hline 
\noalign{\smallskip}
Validity range                 & 60361-- 60377                 &MJD (TDB)\\                      
$t_0$ (Epoch)                  & 60364                         &MJD (TDB)\\
$\nu$                          & 447.871\,561\,130(15)         &Hz       \\

$\dot{\nu}$ & $(3.15\pm 0.36)\times10^{-13}$                       &Hz/s     \\
\chiq{}/d.o.f                  &  74.18/(67-2)=1.14            &         \\
\noalign{\smallskip}
\hline 
\noalign{\smallskip}  
\hline  
\end{tabular}
\label{table:eph} 
}
\end{table} 

\subsection{The outburst light curve} \label{sec:outburst}

From the 1-day binned MAXI \citep{maxi} light curve of \psrtar\ (also presented by \citet{Papitto25}), downloaded from its official website, we can see that the outburst started around MJD 60355, and shows a fast rise within 5 days towards the peak followed by a slow decay during the next $\sim$ 25~days. Therefore, the total `ON' (active) period runs from $\sim$ MJD 60355--60385, and lasts about 30 days.  

The background-subtracted \hxmt\ LE/ME/ HE light curves for the 2--10, 10--35, and 27--60 keV bands dropped during the outburst from $\sim40$ cnt\,s$^{-1}$, $\sim20$ cnt\,s$^{-1}$, and $\sim20$ cnt\,s$^{-1}$, respectively, since the start of the observations to the quiescent level. The outburst showed several reflares during the decay phase of the `ON' state, i.e. the distinct one started at MJD 60365 and peaked at MJD 60367 in the MAXI, \nicer, and \hxmt\ LE/ME data. The outburst profile showed resemblance to those of other AMXPs, such as the recent outbursts from SAX J1808.4--3658 \citep{Illiano23} and IGR J17498--2921 \citep{ZLi24}.

\begin{figure*}[t]
\centering
\includegraphics[width=8.8cm,height=6.5cm]{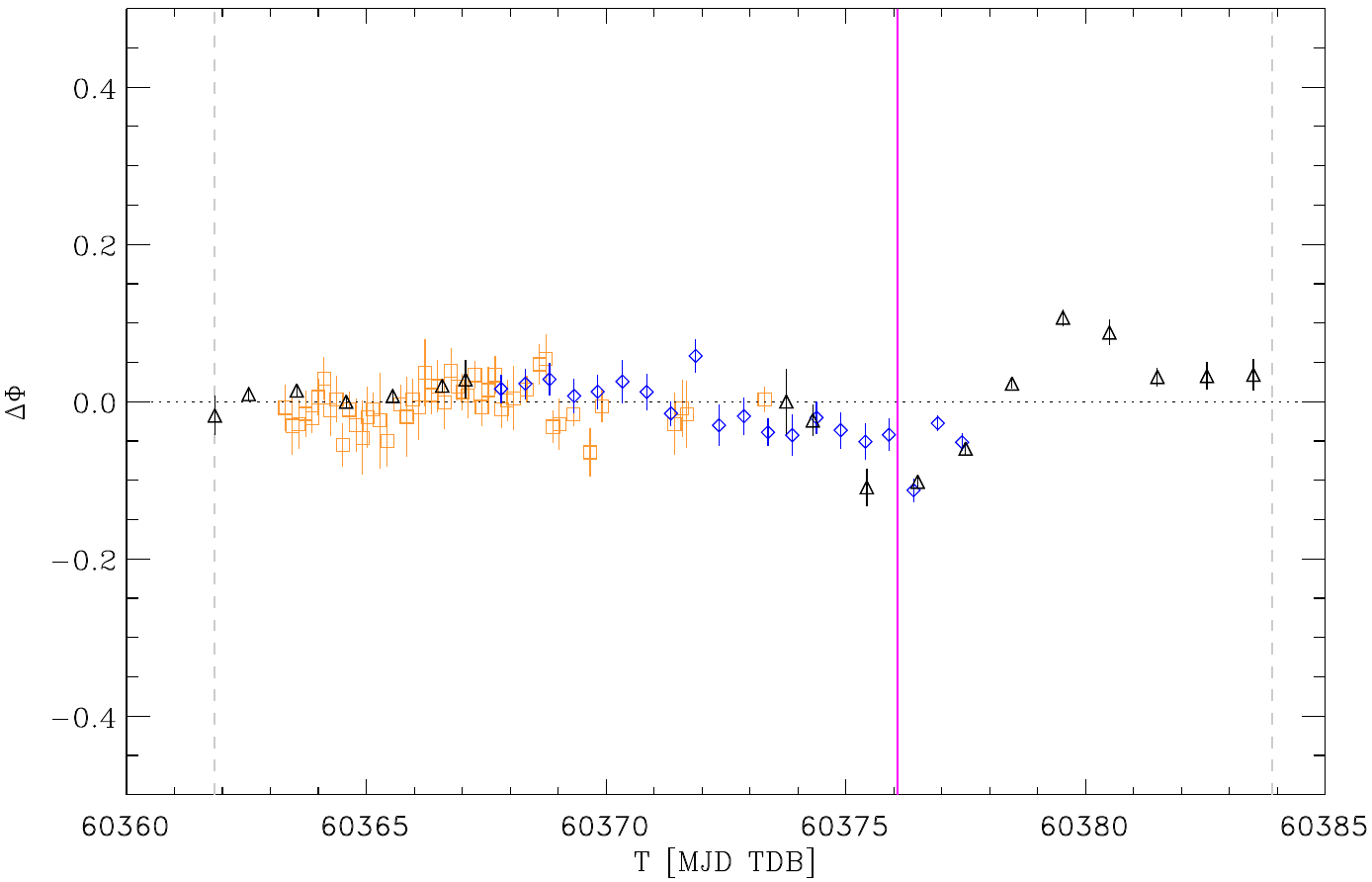}
\includegraphics[width=8.8cm,height=6.5cm]{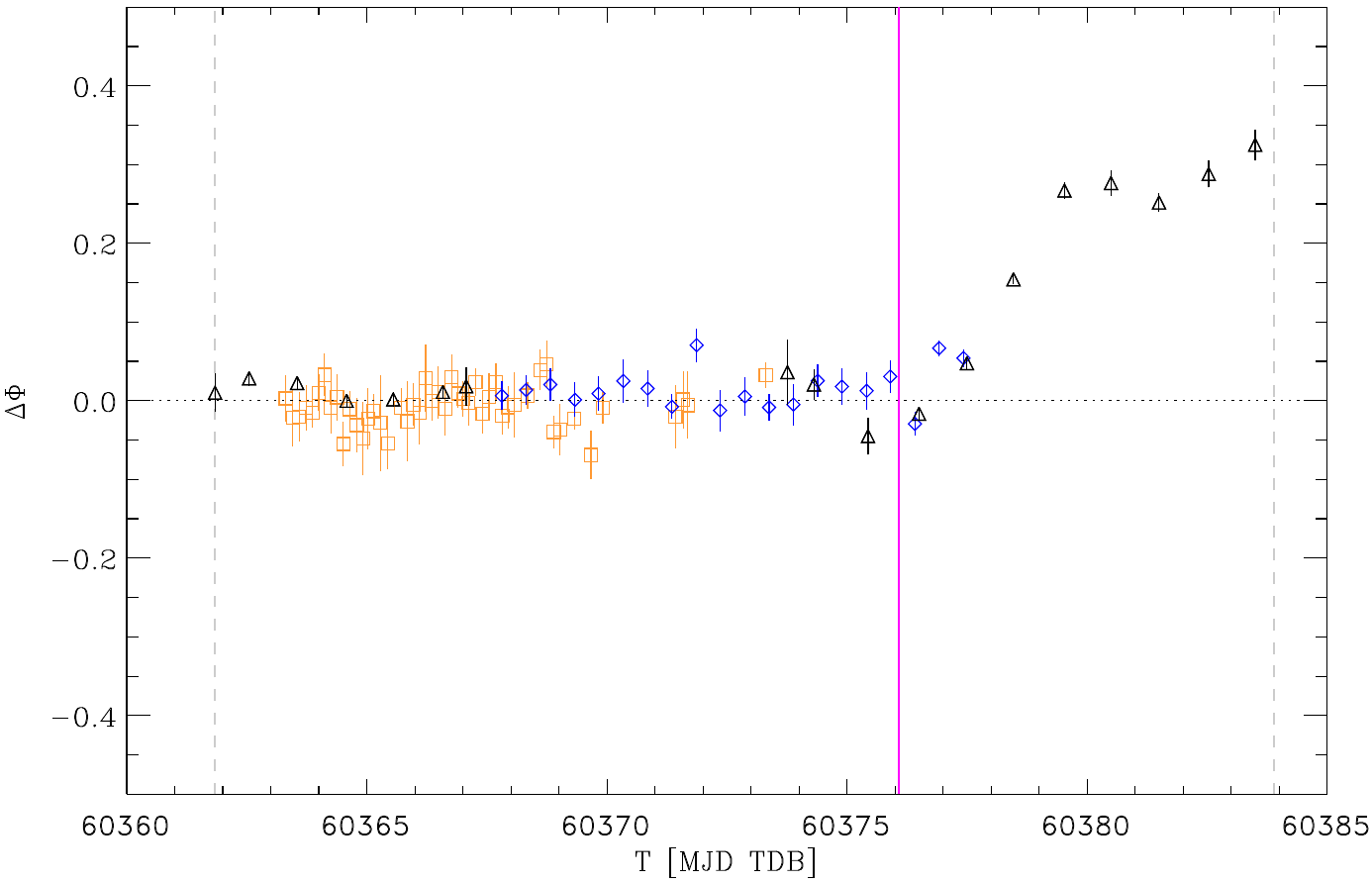}
\caption{The left panel shows the ToA phase-residuals of \nicer\,(black; 3--10 keV), \hxmt-ME\,(orange; 5--10 keV) and \ixpe\,(blue; 3--10 keV) measurements after folding on the 4d-{\tt SIMPLEX} orbital- and timing model (see upper part of Table \ref{table:eph}). This model is based on {\em solely} \nicer\ observations covering the full `ON' outburst phase, from MJD 60361.83 to 60383.89, the boundaries of which are indicated as vertical grey dashed lines. It is evident that while the model keeps phase-coherence systematic un-modeled structures exist. In particular, a swing appeared near MJD 60376.0 (vertical purple line) indicating very likely a change in the accretion process. After this instant highly-increased pulsed emission is detected simultaneously in both \nicer\ and \ixpe\ data. Before the swing - during MJD 60361.83-60376.5 - a curvature trend (downwards) is visible which can be interpreted as a manifestation of a spin-up episode. The spin-up ($\nu, \dot{\nu}$) model (see lower part of Table \ref{table:eph}), derived using \nicer, \hxmt-ME\, and \ixpe\ ToAs across the MJD 60361.83--60377 time interval, is favored against the constant ($\nu$) frequency model (see middle part of Table \ref{table:eph}) at a $8.7\sigma$ level. The pulse-phase residuals of all ToAs collected during the 'ON' phase applying the spin-up model are shown in the right panel.}
\label{fig:toa_residuals}
\end{figure*}

\section{Timing analysis} 
\label{sec:timing}

We performed timing analyses for \nicer, \ixpe, \hxmt, \nustar, \Integ-ISGRI and \epfull-FXT data, covering the energy range of $\sim 1-150$ keV. We ignored (particle) flaring episodes irrespective of the instrument from further analysis. Due to the multitude of X-ray bursts from \psrtar, we excluded the time intervals during which X-ray bursts occurred, to obtain an accurate timing baseline model (see Sect. \ref{sec:tmmod}) for the persistent emission. This baseline model is subsequently used not only to study the energy dependency of the morphology of the pulse-profiles from the persistent emission (see Sect. \ref{sec:tmper}), but also to investigate the pulse profile evolution during the bursts (see Sect. \ref{sec:tmbur}) through separation into pre-burst, burst, and post-burst episodes.  

Using the most accurate source location for \psrtar, as derived from ATCA radio observations \citep{2024ATel16511}, we barycentered the event arrival times, registered at the spacecraft, {of \nicer, \nustar, \ixpe, \Integ\ and \ep-FXT}  adopting the JPL Solar-System DE-405 ephemeris taking into account the instantaneous location of the spacecraft along its orbit around Earth.

\subsection{Timing-model(s) for the persistent emission} \label{sec:tmmod}
Initially, we used the \nicer\ set of monitoring observations covering the `ON' period of the 2024 outburst from MJD 60361.83 to 60383.89 because this provided the most uniform and sensitive exposure to \psrtar\ to construct an accurate timing model describing both the spin of the milli-second pulsar and its orbit around its companion.

\begin{figure*}[t]
\centering
\includegraphics[width=18cm]{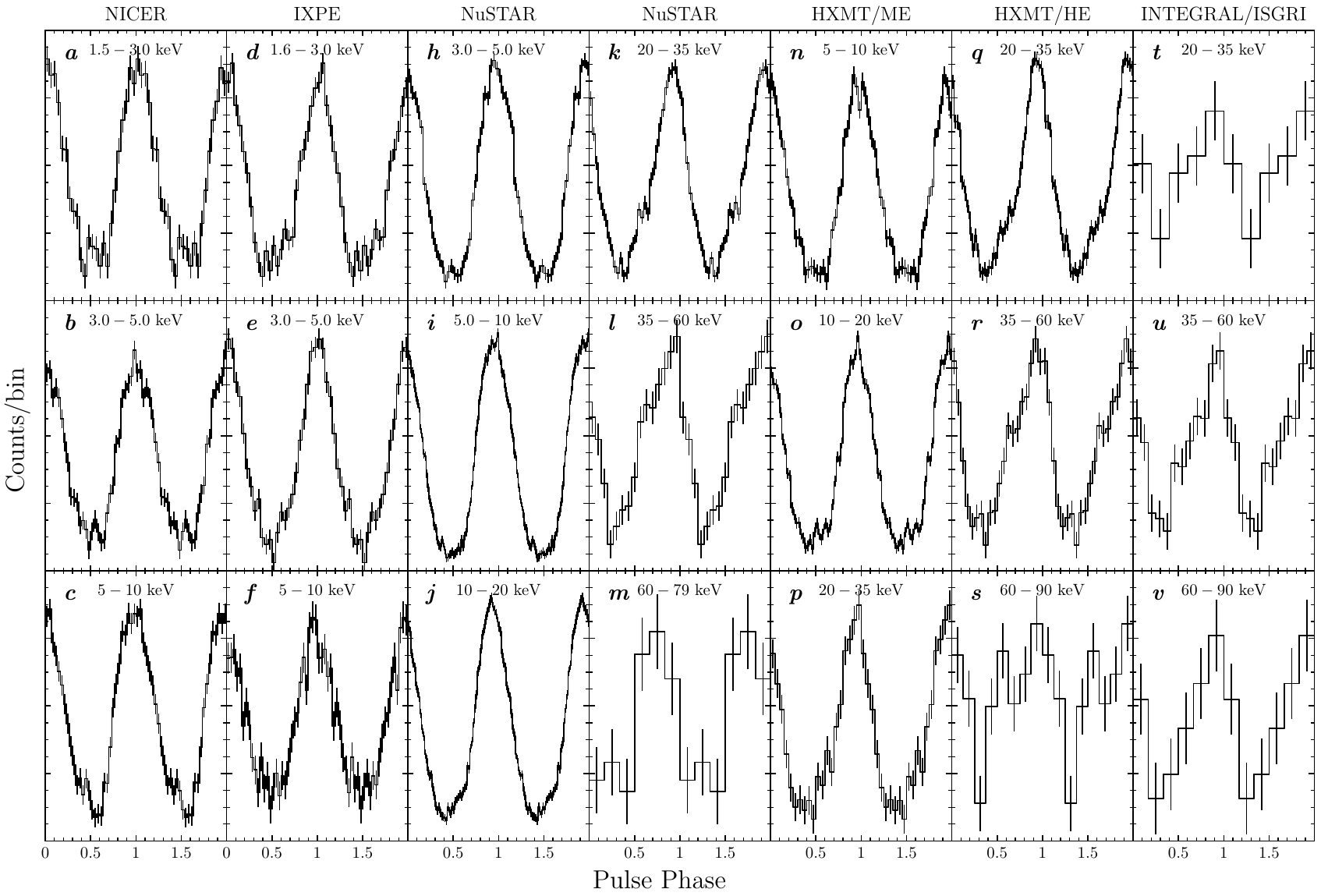}
\caption{The 1.5--90  keV broadband pulse-phase distributions of the persistent emission from \psrtar\ as observed by \nicer\ (panels $\textbf{\textit{a}}$--$\textbf{\textit{c}}$, $1.5-10$ keV), \ixpe\ (panels $\textbf{\textit{d}}$--$\textbf{\textit{f}}$, $1.6-10$ keV), \nustar\ (panels $\textbf{\textit{h}}$--$\textbf{\textit{m}}$, $3-79$ keV), \hxmt\ (panels $\textbf{\textit{n}}$--$\textbf{\textit{p}}$, $5-35$ keV for ME; panels $\textbf{\textit{q}}$--$\textbf{\textit{s}}$, $20-90$ keV for HE), and \Integ/ISGRI (panels $\textbf{\textit{t}}$--$\textbf{\textit{v}}$, $20-90$ keV). The data were taken from observations before the `swing' at $\sim$ MJD 60377.0 applying the `spin-up' model. Two cycles are shown for clarity, while the error bars represent $1\sigma$ errors.  Profiles for energies below $\sim 20$ keV reach their maximum near phase $\sim1$, while above a leading shoulder appears.}
\label{fig:pulse_profile}
\end{figure*}

We used the bin-free $Z_{2}^2$-test statistics \citep{buccheri1983} to evaluate the pulsed signal strength, which is a function of four parameters assuming a constant spin rate of the neutron star and a circular orbit (eccentricity $e\equiv0$).
We employ a 4d optimization scheme based on a downhill {\tt SIMPLEX}\footnote{The downhill {\tt SIMPLEX} method is an optimisation algorithm to find the global minimum of a multi-parameter function. In our case to find the global minimum of the $-Z_1^2$-test statistic, and so to obtain the maximum of the $Z_1^2$ distribution.} algorithm by iteratively improving the $Z_2^2$ statistics with respect to four parameters: the spin frequency $\nu$, the projected semi-major axis of the neutron star $a_{\rm x}\sin i$, the orbital period $ P_{\rm orb} $ and the time-of-ascending node $T_{\rm asc}$ \citep[see e.g.][and references therein for earlier (lower) dimensional versions of the method]{ZLi24}. The best model parameters from this 4d-{\tt SIMPLEX} method are listed in the upper part of Table \ref{table:eph}. The derived values are consistent with those derived by \citet{Ng24} who used a smaller \nicer\ observation set. 

It is noteworthy that, as already noted by \citet{Ng24}, the obtained parameters are significantly different from the model parameters obtained by \cite{2024ATel16480} and \cite{2024ATel16548} who used both smaller \nicer\ and \hxmt-ME observation sets, respectively. The mismatch could be traced back to a convergence onto a secondary (beat) frequency maximum, $\nu_s + \nu_{\hbox{\scriptsize ISS}}$, composed of the true pulsar spin frequency $\nu_s$ and the orbital frequency $\nu_{\hbox{\scriptsize ISS}}$ of the ISS.

We compared the orbital and spin parameters derived in this work (see Table \ref{table:eph}) with those determined by \citet{Molkov24}. Their results are obtained from two SRG/ART-XC observations performed during MJD 60364.333--60367.804 and thus extending over a 3.47-day interval covering about 16 orbital cycle, using a non-well-calibrated onboard clock. Their reported spin-frequency of 447.8718(2) Hz and time-of-ascending node of 60361.64126(5) MJD are consistent with our measurements within $2\sigma$, after accounting for a one orbital cycle difference. However, their derived orbital period of $18804.9(4)$ s and projected semi-major axis of $0.6513(2)$ lt-s lie well beyond the mutual $3\sigma$ uncertainty margins. We attribute this discrepancy to their incoherent timing model approach, which was imposed by unmodeled onboard clock drifts and the limited number of orbital cycles (in our work $\sim 78$ cycles for the spin-up episode (see later) and $\sim 101$ cycles for the full \nicer\ ON period).

\citet{Papitto25} also reported spin- and orbital parameters of \psrtar\ derived from individual analyses of timing data obtained from (a smaller set of) \nicer\  (ignoring data taken between MJD 60367--60383), \ixpe, \xmm and \nustar\ observations. A comparison of the orbital parameters derived by \citet{Papitto25} and those derived in this work shows that these are consistent within the mutual $2\sigma$ uncertainty margins. However, comparing the spin parameters determined by \citet{Papitto25} using \nicer\ observations, providing the longest baseline, with our model(s) (see Table \ref{table:eph}) shows that the models are not consistent and differ by more than $3\sigma$. The cause of this discrepancy could be located in the omission of several \nicer\ observations in the MJD 60367--60383 time interval in the work of \citet{Papitto25} containing important information about the spin evolution. Their \ixpe\ based spin model, however, is consistent with our model(s) within the mutual $2\sigma$ timing margins.

Equipped with accurate orbital parameters (see Table \ref{table:eph}) we can correct for the orbital motion induced periodic variations and derive pulse-arrival times for the various observations performed by the different instruments by applying a Time-of-Arrival (ToA) analysis \citep[see e.g.][for more details]{kuiper2009}. 

In order to avoid the inclusion of energy-dependent phase-shifts, a phenomenon often seen in AMXPs \citep[see e.g.][for IGR J17511-3057, IGR J17591-2342 and IGR J17498-2921, respectively]{falanga11,Kuiper20,ZLi24}, we used compatible energy intervals to derive the ToAs in the event selection process for those instruments with overlapping bandpasses: 3--10 keV for \nicer, 5--10 keV for \hxmt-ME, and 3--10 keV for \ixpe. 
The pulse-arrival time residuals (\nicer; black symbols, \hxmt-ME; orange, and \ixpe; blue) with respect to the 4d-{\tt SIMPLEX} model are shown in the left panel of Fig. \ref{fig:toa_residuals}. It is clear that during the first $\sim$14 days of the outburst the ToAs from the three different instruments nicely overlap and scatter around zero with a slight curvature (downwards) trend until $\sim$ MJD 60376/60377 (vertical purple line) at which a swing occurred indicating likely a change in the accretion process.
After this instant highly increased pulsed emission is detected both for \nicer\ and \ixpe\ \citep[for the latter instrument, see Sect. 3.1 of][]{Papitto25} as indicated by $Z_1^2$-test statistics significances of individual observations.
To quantify the increase in the pulsed flux we estimated the pulsed fraction $P_f$, defined as the ratio of the number of pulsed counts $N_{pul}$ and the number of total counts $N_{tot}$, in the 3--10 keV \nicer\ band for individual observations before and after the swing. For the three \nicer\ observations performed before the swing during MJD 60373--60375 we found, when combined (6.301 ks exposure in total) a $14.3\sigma$ $Z_1^2$ signal and a $P_f$ of $0.059\pm0.012$. The eight \nicer\ observations performed after the swing yielded a $P_f$ in the range 0.13(3) and 0.25(2) with a weighted averaged of $0.178\pm0.008$. This indicates a factor of $\sim 3$ increase of the pulsed fraction crossing the swing, consistent with the findings of \citet{Papitto25} for \ixpe\ data.

The curvature trend before the swing in the ToA residuals (see left panel of Fig. \ref{fig:toa_residuals}) can be interpreted as a manifestation of a spin-up episode. If we fit the MJD 60361.83--60377 \nicer, \hxmt-ME and \ixpe\ ToA data with a spin-up ($\nu, \dot{\nu}$) model, resulting in a spin-up rate of $\dot{\nu}=(3.15\pm0.36)\times 10^{-13}$ Hz/s, this model is favored against a constant ($\nu$) frequency model at a $8.7\sigma$ level applying a maximum likelihood ratio test. The model parameters for both fits are shown in the middle and bottom parts of Table \ref{table:eph}, while the right panel of Fig. \ref{fig:toa_residuals} depicts the ToA residuals of all measurements obtained during the `ON' phase applying the spin-up model. After the swing the pulse arrives progressively later with respect to the spin-up model, indicating likely an evolution to a state of constant spin or even to a spin-down state.

\begin{figure}
\centering
\includegraphics[width=8.5cm]{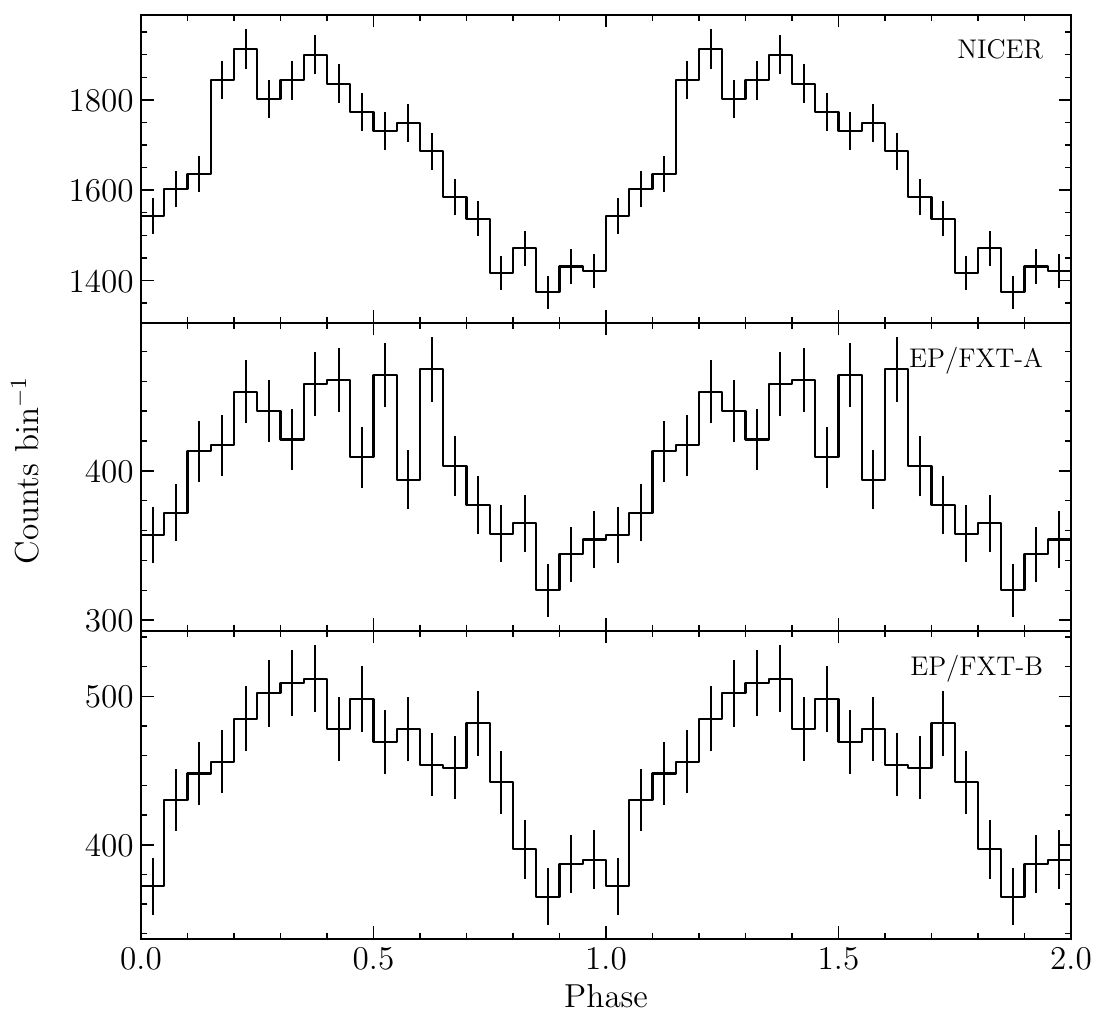}
\caption{The folded pulse profiles (20 bins) of simultaneous observations of \nicer\ (Obs. ids.
6639080112 and 6639080113, top panel) and \ep-FXT A/B (MJD 60382.36--60383.06, middle and bottom panels, respectively) in the 2.5--10 keV band.  }
\label{fig:nicer_fxt}
\end{figure}

It is interesting to note that the \epfull\ observed \psrtar\ between MJD 60382.36 and 60383.06, near the end of `ON' episode, falling between the last two \nicer\ observations (Obs. ids. 6639080112 and 6639080113). 
Data from EP-FXT A/B and NICER were folded in the 2.5--10\,keV range (to mitigate energy-dependent shifts) using the 4d-{\tt SIMPLEX} constant frequency model (Table~\ref{table:eph}). We utilized these simultaneous observations to assess the \epfull's absolute timing accuracy via pulse profile cross-correlation. The phase shift ($\delta\phi$) between EP-FXT A and B was $0.031 \pm 0.039$ ($69 \pm 87\,\mu$s), while between \ep-FXT A and NICER it was $-0.023 \pm 0.018$ ($-51 \pm 40\,\mu$s), suggesting a consistent absolute timing accuracy of \ep-FXT A. However, the phase shift between \ep-FXT B and NICER was $-0.069 \pm 0.028$, indicating a delay of $154 \pm 63\,\mu$s ($2.4 \sigma$ deviation) in the \ep-FXT B pulse arrival relative to NICER, see Fig.~\ref{fig:nicer_fxt}.

\subsection{The persistent emission pulse profiles from \nicer, \ixpe, \nustar, \hxmt\ and \Integ}\label{sec:tmper}
Because the large majority of the data has been taken during the spin-up episode from MJD 60361.83 till $\sim$ MJD 60377 we phase-folded the orbital motion corrected barycentered time stamps of the selected events for all involved instruments upon the `spin-up' timing model to obtain pulse-phase distributions (pulse-profiles) across an as-wide-as possible energy range. This enabled us to derive the lower- and upper bounds of the energy bandpass for which  pulsed emission can be detected as well as to investigate possible morphology changes of the pulse-profile as a function of energy. In Fig. \ref{fig:pulse_profile} the pulse-profiles are shown for the persistent (i.e. non-burst) emission of \psrtar\ from $\sim 1.5$ keV to $\sim 90$ keV using data from \nicer, \ixpe, \nustar, \hxmt-ME/HE and \Integ\ observations. 
\textbf{Above $\sim 90$ keV no significant pulsed emission can be detected. Below 1.5 keV \nicer\ only weakly detects pulsed emission ($6.7\sigma$ applying a $Z_1^2$ test) in the 1-1.5 keV band, while no significant pulsation is found below 1.0 keV.}
From this light curve compilation it is also clear 
that for energies below $\sim$ 20 keV maximum emission is reached near phase 1, while above a leading shoulder (secondary pulse) appears shifting the pulse-averaged emission $\overline{\phi}$ towards earlier phases\footnote{$\overline{\phi}=\int_0^1\phi\cdot F_{pul}(\phi)\cdot d\phi$, in which $F_{pul}$ represents the background subtracted pulse-phase distribution}.

We quantified this shift in detail using \nustar\ data. In particular, we cross-correlated the 5--10, 10--20, 20--35, 35--60 and 60--79 keV pulse profiles with the 3--5 keV profile used as baseline. We obtained the following values (in phase units) for the five energy bands mentioned above, respectively: -0.015(3), -0.037(3), -0.074(4), -0.137(12) and -0.226(34), clearly showing that the higher the energy the earlier the pulse-averaged emission arrives.

\subsection{The pulse profiles of burst, pre-burst, and post-burst epochs}
\label{sec:tmbur}

We investigated potential pulse profile variations affected by X-ray bursts.  We extracted the light curves and events from \hxmt\ ME and HE without filtering good time intervals to avoid missing bursts. We identified 60 X-ray bursts in the ME light curves, of which 40 occurred during periods of low instrumental background \citep[see also][]{Fu25}. For the \ixpe\ and \nustar\ observations, we identified the time intervals of all 52 and 23 X-ray bursts, respectively. The \ixpe, \nustar, and \hxmt\ bursts were studied independently.  

Our analysis focused on specific energy bands: 2--8\,keV for \ixpe, 3--10\,keV, 10--35\,keV, and 35--60\,keV for \nustar, 5--30\,keV for \hxmt/ME, and 20--60\,keV for \hxmt/HE. For each instrument, we determined the burst peak time as the reference ($t = 0$) in the
following analysis \citep[see e.g.,][]{Ji24}. Then, burst start time, $t_{\rm start}$, and stop time,  $t_{\rm stop}$, were defined as $t-15$ s and $t+35$ s, respectively. Due to the limited number of photons in individual bursts, we stacked the data from all identified bursts within the $[t_{\rm start}, t_{\rm stop}]$ intervals for each instrument and energy band to obtain sufficient signal-to-noise for pulse profile analysis. For comparison with the persistent emission, we defined pre-burst intervals as $[t_{\rm start} - 200\,\mathrm{s}, t_{\rm start} - 50\,\mathrm{s}]$ and post-burst intervals as $[t_{\rm stop} + 50\,\mathrm{s}, t_{\rm stop} + 200\,\mathrm{s}]$, relative to each burst. Data from these 150\,s pre- and post-burst epochs were similarly stacked. 

\begin{figure} 
\centering
\includegraphics[width=\linewidth]{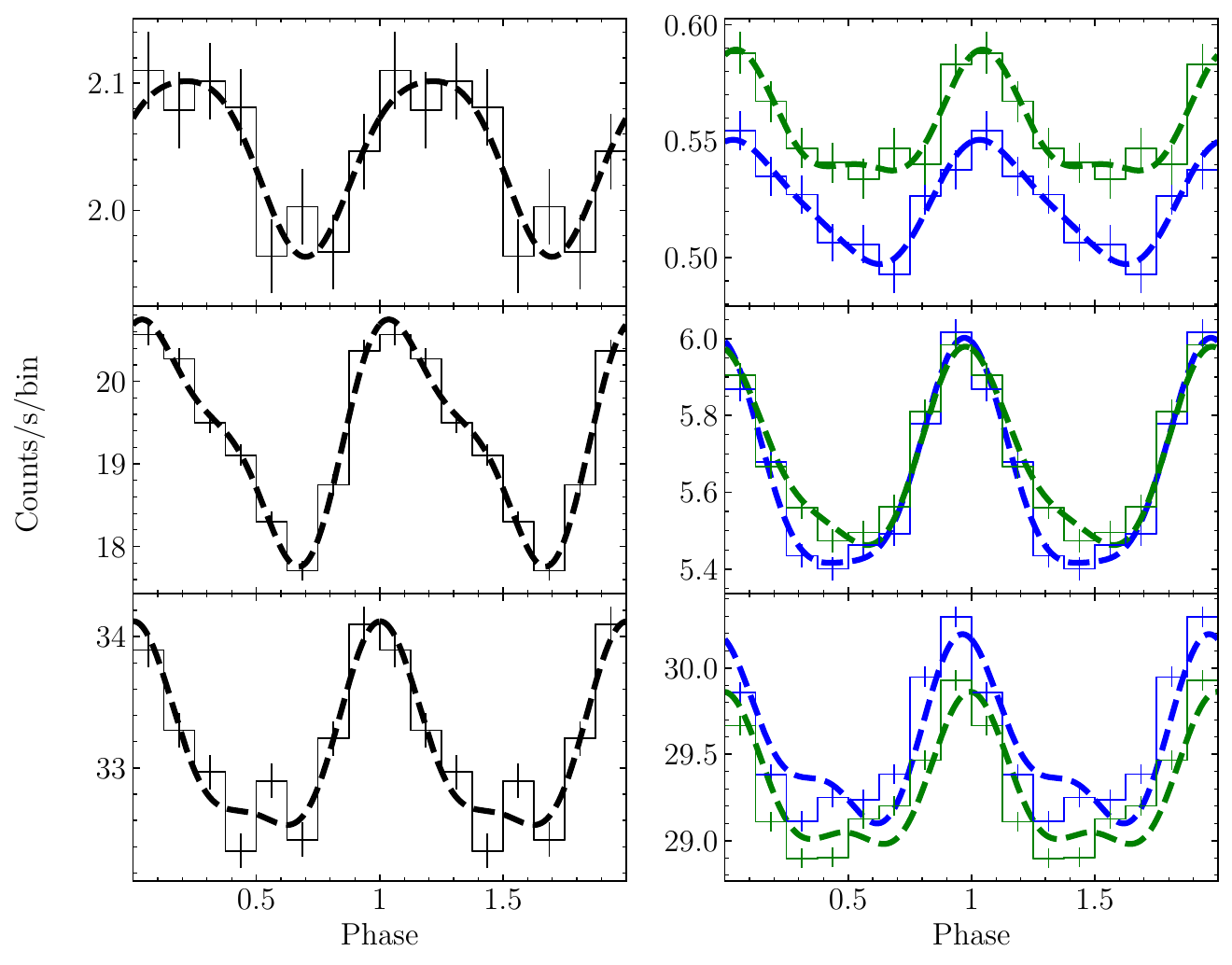}
\caption{The burst, pre-burst, and post-burst pulse profiles from \hxmt\ and \ixpe. Left panels show the burst profiles with 8 bins from \ixpe\ in 2--8 keV (top panel), \hxmt/ME 5--30 keV (middle panel), and \hxmt/HE 20--60 keV (bottom panel). Right panels show the pre-burst (blue) and post-burst (green) pulse profiles, with the same energy band and instrument as the  left panels.   Vertical error bars indicate 1$\sigma$ uncertainties. The best-fitted Fourier series by using Equation~\ref{equ:four} are shown for each profile.
}
\label{fig:IXPE_HXMT_profile}
\end{figure}

\begin{figure} 
\centering
\includegraphics[width=\linewidth]{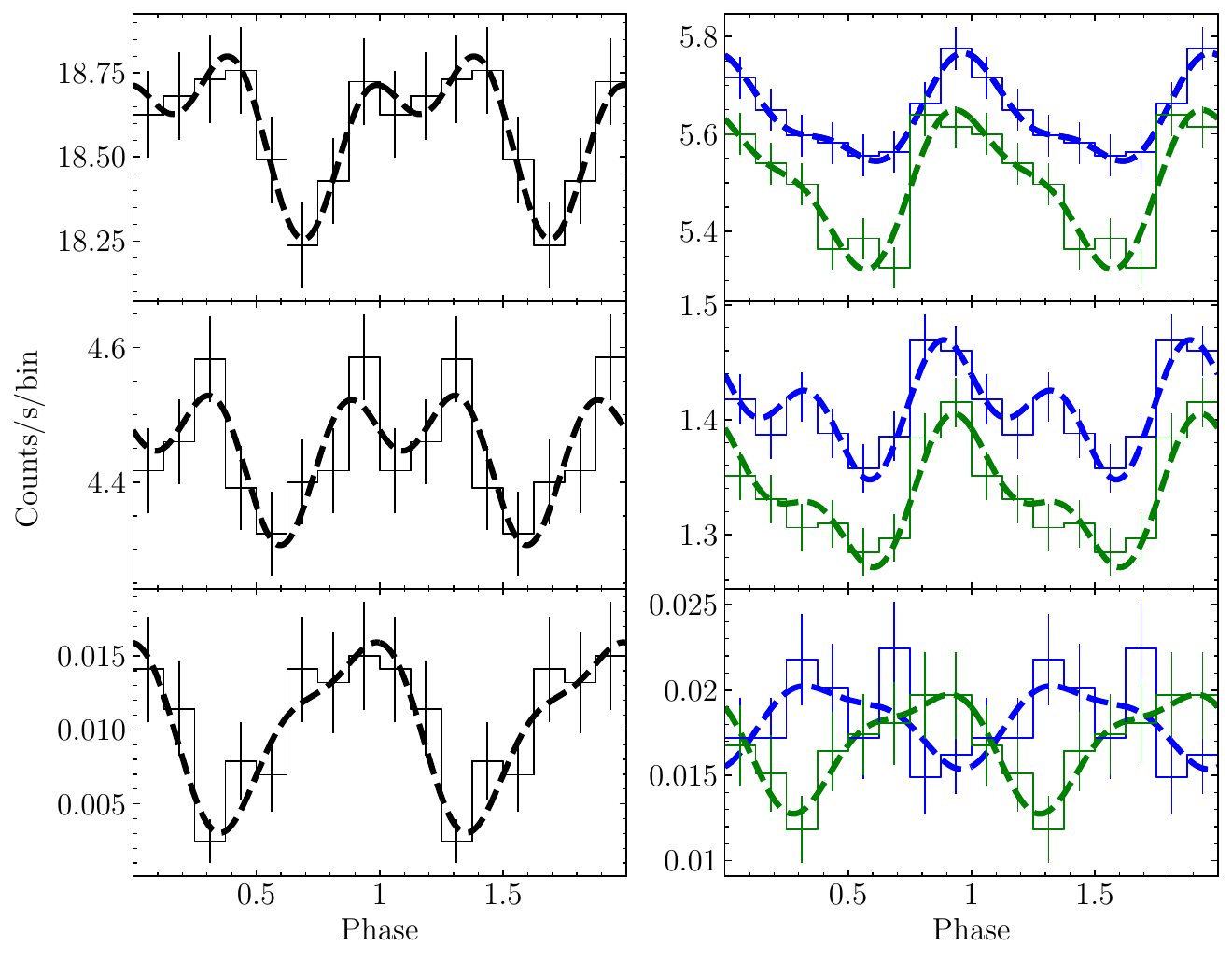}
\caption{The burst, pre-burst, and post-burst pulse profiles from \nustar. Left panels show the burst profiles with 8 bins from \nustar\ in 3--10 keV (top panel), 10--35 keV (middle panel), and  35--60 keV (bottom panel). Right panels show the pre-burst (blue) and post-burst (green) pulse profiles, with the same energy band as the  left panels.   Vertical error bars indicate 1$\sigma$ uncertainties. The best-fitted Fourier series by using Equation~\ref{equ:four} are shown for each profile.} %
\label{fig:Nustar_profile}
\end{figure}

Pulse profiles for the stacked burst, pre-burst, and post-burst epochs were generated by folding the corresponding event data using the orbital and spin ephemeris presented in Table~\ref{table:eph}. Significant pulsations were detected in all three epochs (burst, pre-burst, post-burst) across the selected energy bands. The folded profiles, corrected for exposure time, are shown in Figs.~\ref{fig:IXPE_HXMT_profile} (\ixpe\ and \hxmt/ME/HE) and \ref{fig:Nustar_profile} (\nustar).

We fitted the pulse profiles with a truncated Fourier series given by a formula
\begin{equation}\label{equ:four}
F(\phi) = A_0 + \sum_{k=1}^2 A_k \ \cos[2\pi\ k (\phi-\phi_k)],
\end{equation}
where $A_0$ is the constant level of the profile,  $A_{1}$ and $A_{2}$ are the amplitudes,  $\phi_1$ and $\phi_2$ are the phase angles, of the fundamental and the first overtone, respectively.

For the \ixpe\ data (2--8\,keV), the pre- and post-burst profiles exhibit consistent shapes (similar relative harmonic amplitudes $A_k/A_0$ and phases $\phi_k$), differing primarily in normalization ($A_0$), with the post-burst rate ($0.56 \pm 0.01$~cnt~s$^{-1}$) being slightly higher than the pre-burst rate ($0.52 \pm 0.01$~cnt~s$^{-1}$; see top panels of Fig.~\ref{fig:IXPE_HXMT_profile}). The stacked burst profile ($A_0 = 2.04 \pm 0.01$~cnt~s$^{-1}$), however, shows a significantly larger fundamental amplitude ($A_1 = 0.069 \pm 0.020$~cnt~s$^{-1}$, compared to $\approx 0.025 \pm 0.005$~cnt~s$^{-1}$ for pre/post burst) and displays a phase lag of the fundamental component, $\Delta\phi_1 \approx 0.15$, relative to the average pre-/post-burst phase. Notably, the fractional amplitude $A_1/A_0$ decreases during the burst ($0.03 \pm 0.01$) compared to the persistent emission ($0.04$--$0.05 \pm 0.01$). The first overtone ($A_2$) is weak or insignificant in all epochs for \ixpe.

In the \hxmt/ME band (5--30\,keV), the pre- and post-burst profiles are statistically consistent in both shape and normalization ($A_0 \approx 5.6$--$5.7$~cnt~s$^{-1}$; middle panels of Fig.~\ref{fig:IXPE_HXMT_profile}). The stacked burst profile is markedly different: the normalization ($A_0 = 19.32 \pm 0.05$~cnt~s$^{-1}$) and harmonic amplitudes ($A_1 = 1.31 \pm 0.08$~cnt~s$^{-1}$, $A_2 = 0.41 \pm 0.07$~cnt~s$^{-1}$) are $4\sim6$ times larger than in the persistent emission ($A_1 \approx 0.25$--$0.29$, $A_2 \approx 0.06$--$0.07$~cnt~s$^{-1}$) due to the strong burst contribution. The fractional amplitude $A_1/A_0$ is slightly higher during the burst ($0.065 \pm 0.001$) compared to persistent emission ($0.046$--$0.051 \pm 0.001$). The first overtone ($A_2$) is clearly significant in the burst profile ($A_2/A_0 = 0.021 \pm 0.001$). Furthermore, the fundamental phase during bursts shows a significant lag of $\Delta\phi_1 \approx 0.11$ compared to the persistent emission phase, corresponding to a time delay of $\approx 0.25$\,ms.

For the \hxmt/HE data (20--60\,keV), the burst profile amplitudes are only marginally higher than the pre-/post-burst profiles (e.g., $A_0$ increases from $\approx 29.4$ to $33.15 \pm 0.10$~cnt~s$^{-1}$, $A_1$ from $\approx 0.44$ to $0.74 \pm 0.15$~cnt~s$^{-1}$; bottom panels of Fig.~\ref{fig:IXPE_HXMT_profile}), indicating a smaller relative contribution from the burst flux in this harder band compared to the 5--30\,keV band. The fractional amplitude $A_1/A_0$ is comparable during the burst ($0.020 \pm 0.001$) and persistent emission ($0.016$--$0.018 \pm 0.001$). A phase lag in the fundamental is still detected relative to the persistent emission, $\Delta\phi_1 \approx 0.02$, corresponding to a time delay of $\approx 0.045$\,ms.

The \nustar\ observations contained fewer bursts, resulting in pulse profiles with larger statistical fluctuations (Fig.~\ref{fig:Nustar_profile}). In the 35--60\,keV band, the pre-burst ($\phi_1 = 0.43 \pm 0.11$) and post-burst ($\phi_1 = 0.81 \pm 0.04$) profiles appear statistically different in phase. As the post-burst profile shape in this band seems more consistent with the overall persistent emission profile (Sect.~\ref{sec:timing}, Fig.~\ref{fig:pulse_profile}), we adopt the post-burst profile phase as the reference for calculating the phase lag in this specific band. The energy-dependent behavior observed by \nustar\ consistent with the other instruments in the similar energy bands. In the 3--10\,keV band, similar to \ixpe, the fractional amplitude $A_1/A_0$ is lower during the burst ($0.01 \pm 0.001$) compared to the persistent emission ($0.017$--$0.03 \pm 0.01$). The phase lag ($\Delta\phi_1$) of the fundamental component during bursts, relative to the persistent emission, decreases systematically with increasing energy, from $\Delta\phi_1 \approx 0.21$ (3--10\,keV) to $\approx 0.10$ (10--35\,keV) and $\approx 0.08$ (35--60\,keV, relative to post-burst).

\section{Broadband spectral analysis} \label{sec:spectra}

\subsection{\nicer\ spectral fitting}\label{sec:nicer_spec}
We fitted the \nicer\ spectra collected between MJD 60361.84-60391.67, using \textsc{xspec} version 12.12.1 \citep{arnaud96}.  All uncertainties of the spectral parameters are provided at a $1\sigma$ confidence level for a single parameter.  We fit all spectra by using the thermally Comptonized continuum, {\tt nthcomp} modified by the interstellar absorption. The {\tt nthcomp} model is defined by an asymptotic power-law photon index, $\Gamma$, and the temperatures of the electron cloud, $kT_{\rm e}$, and seed photons, $kT_{\rm BB}$. We assumed a blackbody seed photons distribution emitted from the NS surface. The absorption was described by the {\tt tbabs} model, for which we adopted the interstellar abundances of \citet{wilms00} and the photoelectric absorption cross sections of \citet{Verner96}. Additionally, we included a Gaussian emission line at $\sim1.7$ keV to model a potential instrumental Si fluorescence line from the Focal Plane Modules \citep[see e.g.,][]{Marino22}.  The full model is {\tt tbabs$\times$(gaussian+nthcomp)} in \textsc{xspec}. 

 \begin{figure} 
\centering
\includegraphics[width=0.95\linewidth]{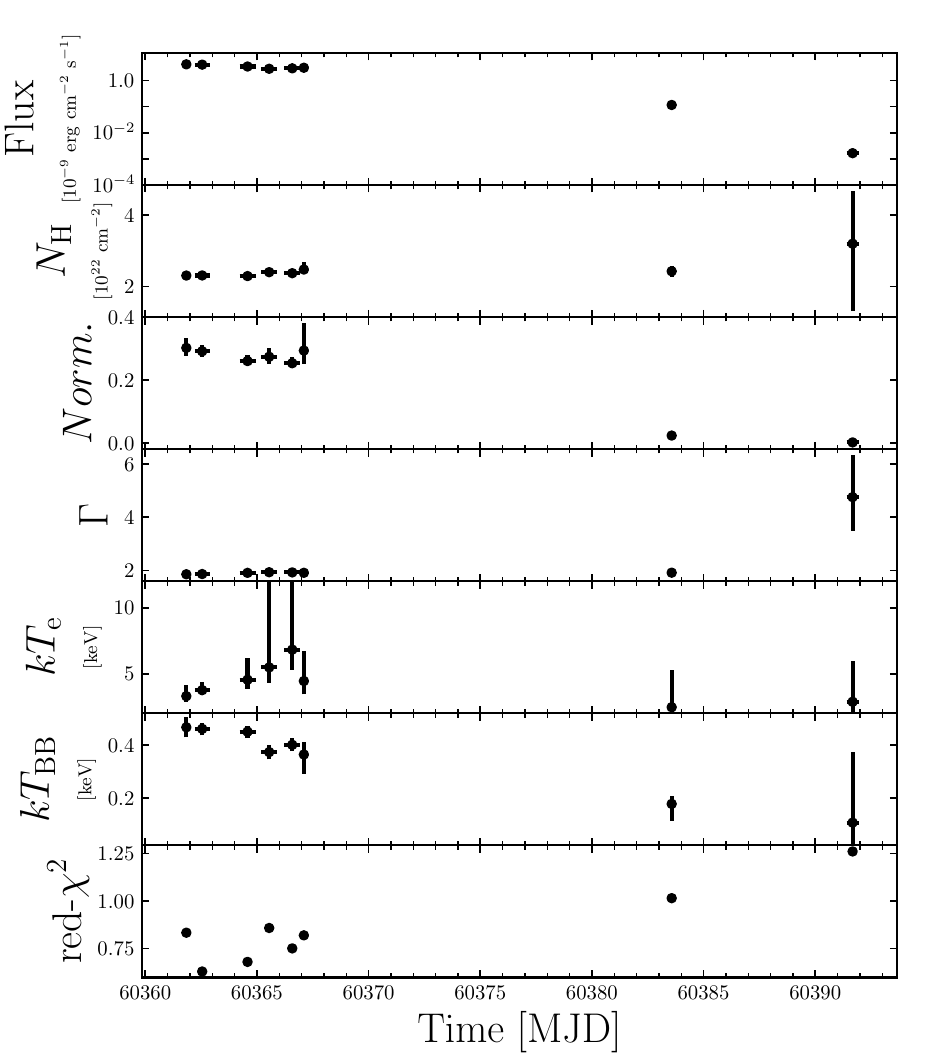}
\caption{The best-fitted parameters of the \nicer\ spectra from \psrtar\ by using the model {\texttt{tbabs$\times$(gaussian+nthcomp)}}. From top to bottom, the bolometric flux, the hydrogen column density, $\Gamma$, the electron temperature, $kT_{\rm e}$, the blackbody temperature, $kT_{\rm BB}$, and the reduced $\chi^2$. The parameters of \texttt{gaussian} are not shown here.
}
\label{fig:nthcomp}
\end{figure}

The best fitted parameters are shown in Fig.~\ref{fig:nthcomp}. The last spectrum has low counting statistics, and the uncertainties of the best-fitted parameters are large. So, we do not report its parameters but only bolometric flux.   The spectra can be well fitted with reduced $\chi^2<1.25$.  We calculated the unabsorbed bolometric flux in the 1--250 keV range using the tool \texttt{cflux}, which is used to estimated the average accretion rate in Sect.~\ref{sec:mag}. It is worthy to note that the estimated bolometric flux could be biased due to lacking of observations above 10 keV. Nevertheless, this method provides a consistent way to track the luminosity evolution.    During the outburst,  the disk blackbody temperature decreased from 0.4 to 0.1 keV. The  hydrogen column density did not change much, and the mean value is $(2.4\pm0.1)\times10^{22}~{\rm cm^{-2}}$. The optical depth was in the range 1.7--3.0. The electron temperature was below 7 keV. The photon index $\Gamma$ remained stable at a value of $\sim1.9$, with the exception of the final observation, which was poorly constrained at a value of $\sim4.75$. These parameters of the \texttt{nthcomp} model are broadly consistent with \hxmt\ results \citep{Fu25}.   The bolometric flux dropped from a peak value of $4\times10^{-9}$ to $1.7\times10^{-12}~{\rm erg~cm^{-2}~s^{-1}}$ at MJD 60391.4.

\subsection{Joint \nicer, \nustar, and \hxmt\ spectral fitting}

We performed broadband spectral analysis using quasi-simultaneous observations from \nicer, \nustar, and \hxmt. Specifically, we utilized \nustar\ data obtained between MJD 60366.46--60369.88, contemporaneous with \nicer\ ObsIDs 6639080103 (MJD 60366.22--60366.94) and 6639080104 (MJD 60367.07--60367.14), and \hxmt\ ObsIDs P061437300207--P061437300214 (MJD 60365.91--60366.94) and P061437300301 (MJD 60367.096--60367.101).

To perform joint spectral fitting based on the observation overlaps, we divided the \nustar\ data into two epochs:
\begin{itemize}
    \item \textbf{Epoch 1:} MJD 60366.46--60367.0 (\nustar\ exposure: 23.8 ks). Jointly fitted with \nicer\ ObsID 6639080103 (5.2 ks) and combined \hxmt\ data from ObsIDs P061437300207--P061437300214 (ME: 15.2 ks, HE: 15.4 ks).
    \item \textbf{Epoch 2:} MJD 60367.0--60369.88 (\nustar\ exposure: 131.3 ks). Jointly fitted with \nicer\ ObsID 6639080104 (0.2 ks) and \hxmt\ ObsID P061437300301 (ME: 1.7 ks, HE: 0.3 ks).
\end{itemize}
Initial comparisons revealed significant discrepancies between \hxmt/LE and \nicer\ spectra in the soft X-ray band, which can not be resolved by adding a simple cross-calibration constant. Considering that \nicer\ spectra have higher photon statistic, therefore, the LE data were excluded from the spectral fitting. Based on calibration recommendations and observed data quality, we adopted the following energy ranges for spectral fitting: 1--10 keV for \nicer\ \citep[see e.g.][]{ZLi24}, 4--79 keV for \nustar/FPMA \& FPMB (ignoring data below 4 keV due to persistent residuals observed in joint fits, consistent with findings in other LMXBs, e.g., \citealt{Ludlam20, ZLYu24, Adegoke24}), 8--20 keV for \hxmt/ME, and 30--80 keV for \hxmt/HE \citep{XBLi20}. A systematic error of 1\% was added to the spectra from each instrument to account for potential residual calibration uncertainties.

\begin{figure} 
\centering
\includegraphics[width=0.95\linewidth,angle=0]{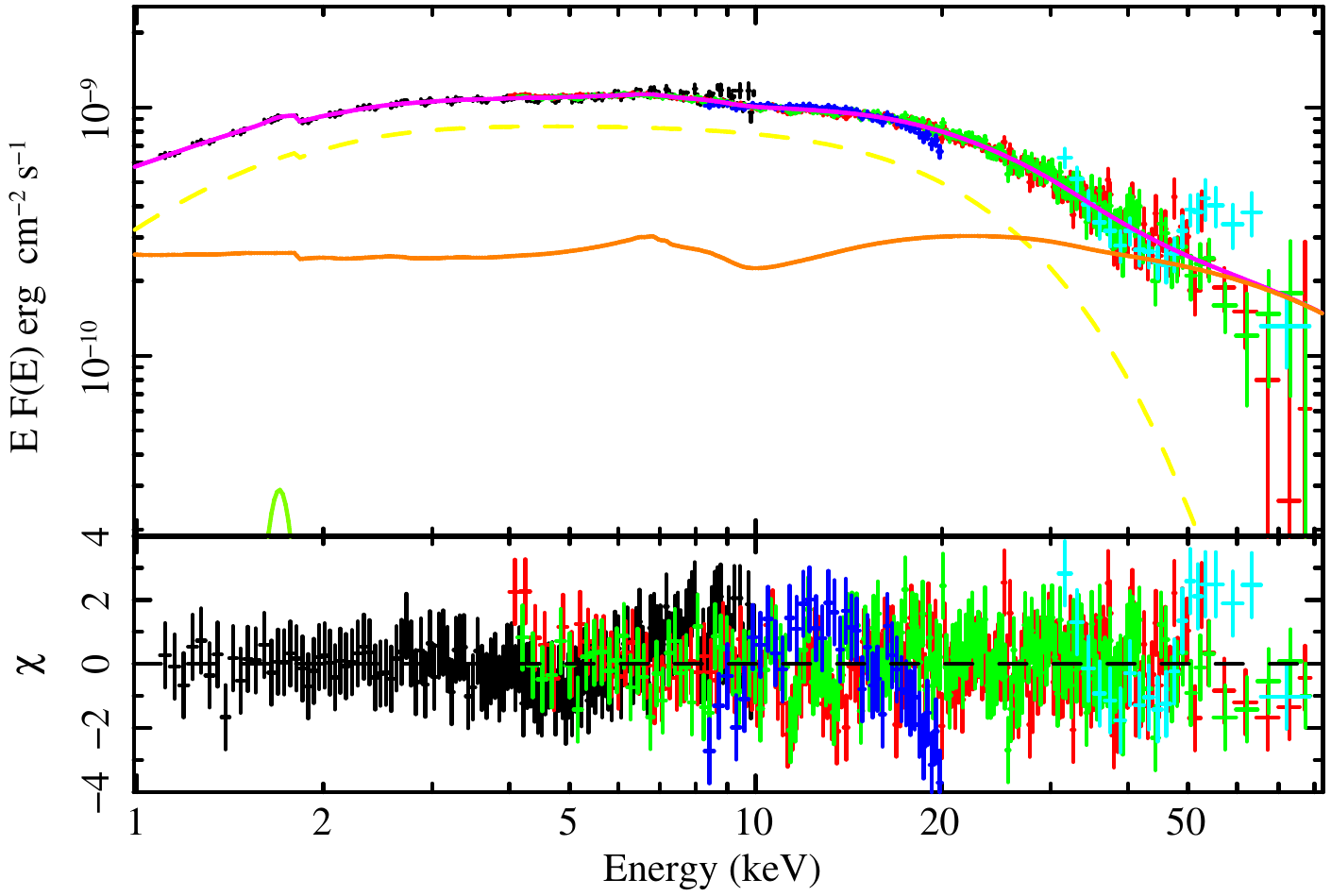}
\includegraphics[width=0.95\linewidth,angle=0]{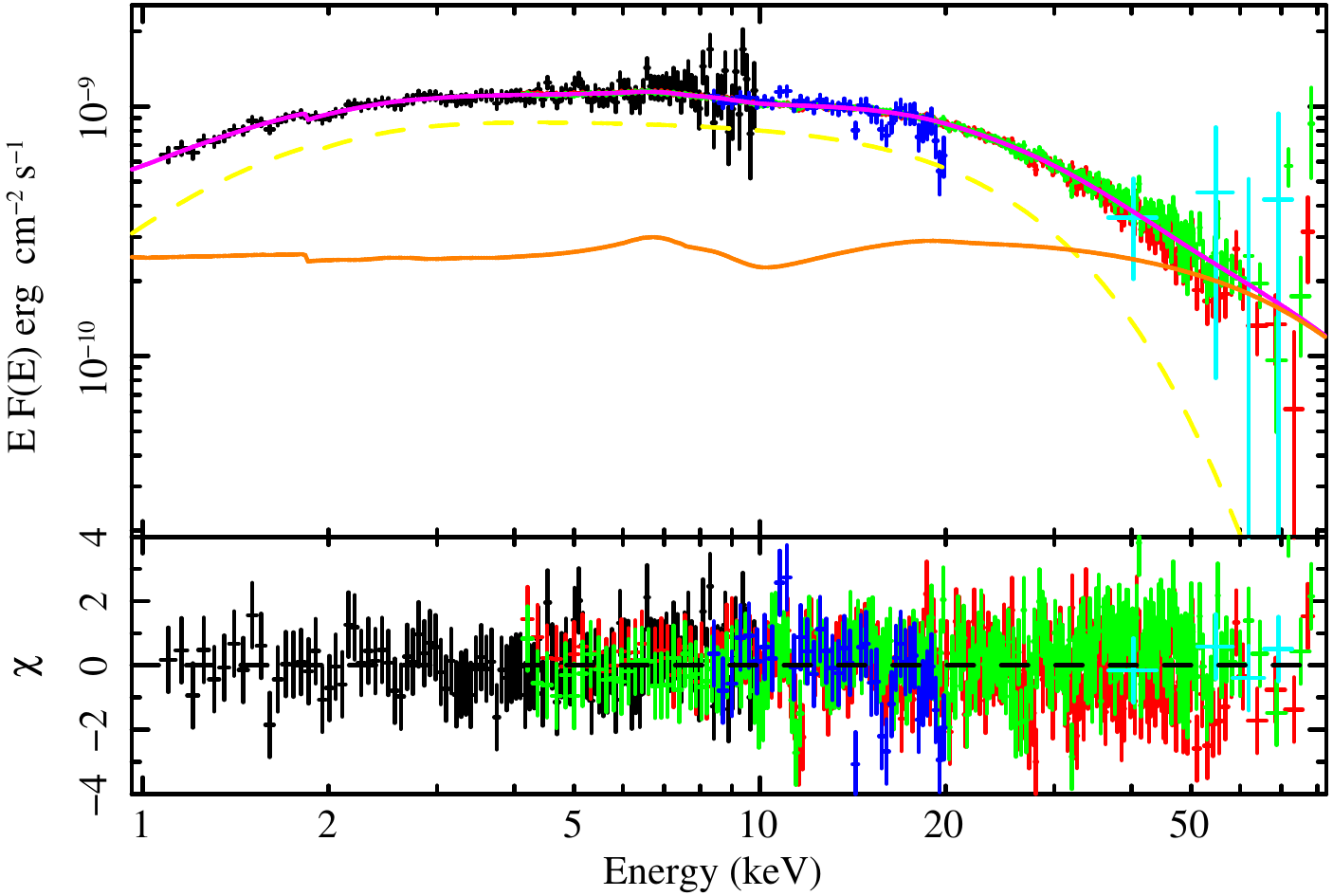}
\caption{Joint \nicer, \nustar, and \hxmt\ spectra fitting from \psrtar\ by using the model {\tt constant$\times$tbabs$\times$(nthcomp+gaussian+relxillCp)$\times$edge}. In the top panel, the spectra are Epoch 1 from \nicer\ (Obs. Id. 6639080103, MJD 60366.22--60366.94, 5.2 ks, black points), \nustar\ (MJD 60366.46--60367.0, 23.8 ks, red and green points), \hxmt\ ME/HE (Obs. Ids. P061437300207--P061437300214, 15 ks, blue and cyan points), respectively. In the bottom panel, the spectra are Epoch 2 from \nicer\ (Obs. Id. 6639080104, MJD 60367.07--60367.14, 0.2 ks) in 1--10 keV, \nustar\ (MJD 60367.0--60369.8, 131.3 ks), \hxmt\ ME/HE (Obs. Id. P0614373003, 1.7 ks for ME and 0.3 ks for HE), respectively. The energy ranges for \nicer\, \nustar\, \hxmt/ME/HE are 1--10, 4--79, 8--20, 30--80 keV, respectively.  Note the \texttt{gaussian} component is included only in Epoch 1. The \texttt{nthcomp}, \texttt{relxillCp}, and the total model are shown as the dashed yellow, solid orange, and solid magenta lines, respectively. 
}
\label{fig:broadband}
\end{figure}

\begin{table}[hbtp]
\begin{ruledtabular}
\centering
 \caption{\label{table:spec} Best-fit spectral parameters  of    the \nicer/\nustar/\hxmt\ data for \psrtar\ using the model  {\tt constant$\times$tbabs$\times$(nthcomp+gaussian+relxillCp)$\times$edge}.}
 \centering
 \begin{tabular}{lll} 
&Epoch 1 & Epoch 2 \\
 \hline 
Parameter (units) &Best-fit values & \\

  \hline 
   & \texttt{tbabs}   &     \\
$N_{\rm H}~(10^{22}~ {\rm cm}^{-2})$& $ 2.63\pm0.05 $  & $  2.60\pm 0.07$ \\
   & \texttt{nthcomp}   &     \\
$\Gamma$                            & $ 1.99 \pm0.01 $      & $  2.02\pm0.02$ \\ 
$kT_{\rm e}$ (keV)                        & $ 5.35 \pm0.11 $      & $  6.18\pm0.13$ \\ 
$kT_{\rm BB}$ (keV)                       & $ 0.40 \pm0.02 $      & $  0.42\pm0.02$ \\ 
Norm$_{\rm nthcomp}$                & $ 0.21 \pm0.02 $      & $  0.20\pm0.02$ \\ 
   & \texttt{gaussian}\textsuperscript{a}   &     \\
$E$  (keV)                       & $ 1.69 \pm0.10 $      & - \\ 
$\sigma$   (keV)                     & $ 0.08 \pm0.07 $      & - \\ 
Norm$_{\rm gaussian}~(\times10^{-3})$      & $ 1 \pm0.02 $      & - \\ 

   & \texttt{relxillCp}   &     \\
$i$ (deg)                           & $31_{-14}^{+11} $  & $  68\pm5$ \\  
$R_{\rm in} (R_{\rm ISCO} \textsuperscript{b})$         & $ 41_{-26}^{+ 37} $  & $ 40_{-21}^{+ 37}$ \\  
$\Gamma$                            & $  1.85 \pm0.04 $      & $  1.88\pm0.04$ \\  
$\log(\xi/{\rm erg~cm~s^{-1}})$     & $ 3.69\pm0.08 $  & $  3.66\pm0.07$ \\  
$\log(n_{\rm e}/{\rm cm^{-3}})$     & $16\pm1 $  & $ 16\pm 1$ \\  
$A_{\rm Fe}$ (solar)                & $ 1.1\pm0.3 $  & $  1.0\pm0.2$ \\  
$kT_{\rm e}$      (keV)             & $225\pm115 $  & $53\pm10$ \\  
$f_{\rm refl.}$                     & $ -0.7\pm0.2 $  & $ -0.9\pm0.2$ \\  
Norm$_{\rm refl.}~(\times10^{-3})$  & $ 2.9\pm0.1$  & $  4.0\pm0.8$ \\  
$E_{\rm Edge} $ (keV)               & $ 1.82\pm0.03 $  & $  1.83\pm0.09$ \\  
$\tau $                             & $ 0.07 \pm0.01 $  & $  0.06\pm0.02$ \\  
   & \texttt{constant}\textsuperscript{c}   &     \\

$C_{\rm \nicer}$              & 1 (fixed)  & 1 (fixed) \\  
$C_{\rm \nustar/FPMA}$              & $ 0.93 \pm0.01 $  & $  0.74\pm0.01$ \\  
$C_{\rm \nustar/FPMB}$              & $ 0.95 \pm0.01 $  & $  0.76\pm0.01$ \\  
$C_{\rm \hxmt/ME}$                    & $ 0.89 \pm0.01 $  & $  0.86\pm0.01$ \\  
$C_{\rm \hxmt/HE}$              & $ 1.14\pm0.04 $  & $  0.99\pm0.38$ \\ 
  \hline 
 $\chi^{2}/{\rm d.o.f.}$ &   688.3/642 &   646.7/644\\ 
 $F_{\rm bol}$ ($10^{-9}$ erg s$^{-1}$ cm$^{-2}$)\textsuperscript{d} & $3.85\pm0.01$ & $3.81\pm0.01$ \\ 
\end{tabular}
\end{ruledtabular}
\begin{tablenotes}
\item[a] \textsuperscript{a} For Epoch 2, by adding a \texttt{gaussian} component with a centroid energy around 1.7 keV only improved the $\chi^2$ by less than 0.1, indicating that it is not necessary for this epoch. 
\item[b] \textsuperscript{b} The radius of the innermost circular orbit, $R_{\rm ISCO}=6GM/c^2$.
\item[c]  \textsuperscript{c} The multiplication factor for all instruments is provided.
\item[d]  \textsuperscript{d} Unabsorbed flux in the 1--250 keV energy range.
\end{tablenotes}
 \label{tab:broadband_spec} 
\end{table}

For the broadband spectra for both epochs, an initial fit with a simple absorbed Comptonization model yielded statistically unacceptable results ($\chi^2/\mathrm{dof} \gg 2$). We therefore adopted a more physically motivated model incorporating relativistic reflection \citep{relxill1,relxill2,relxill3}, which has been successfully applied to other AMXPs \citep[e.g.,][]{ZLi23, ZLi24, Ludlam24}. The final model adopted was \texttt{constant $\times$ tbabs $\times$ (nthcomp + gaussian + relxillCp) $\times$ edge}.  

The free parameters of the reflection model are as follows: the binary inclination, $i$, the inner and outer radius of the disc, $R_{\rm in}$ and $R_{\rm out}$, the power law index of the incident spectrum, $\Gamma$, the electron temperature in the corona, $kT_{\rm e}$, the logarithm of disk ionization, $\log(\xi/{\rm erg~cm~s^{-1}})$, the iron abundance normalized to the Sun, $A_{\rm Fe}$, the density of the disk in logarithmic units, $\log (n_{\rm e}/{\rm cm^{-3}})$, and the reflection fraction, $f_{\rm refl.}$. We fixed the inner and outer emissivity indices, $q_1$ and $q_2$, both at 3 to Newtonian emissivity \citep{Reynolds03}, the break radius between these two emissivity indices and the outer disk radius, $R_{\rm out}=R_{\rm break}=1000R_{\rm g}$, where $R_{\rm g}=GM_{\rm NS}/c^2$ is the gravitational radius, $G$ and $c$ are the gravitational constant and the speed of light, respectively. A negative reflection fraction ($f_{\rm refl.} < 0$) was used to model the reflection component without including the direct illuminating continuum, which is modeled separately by the explicit \texttt{nthcomp} component. For \psrtar\ spinning at 448 Hz, we obtained the dimensionless spin parameter $a = 0.21$ using the relation $a=0.47/P$ where $P$ is the spin period in unit of ms \citep{Braje00}, which was also fixed.  
An emission feature, $\texttt{gaussian}$, required only for Epoch 1 to model a residual around 1.7 keV. Its energy, width ($\sigma$), and normalization were free. The \texttt{constant} accounts for cross-calibration normalization factors and possible flux variations between instruments, which was fixed to 1 for \nicer\ and allowed to vary for \nustar/FPMA, FPMB, \hxmt/ME, and \hxmt/HE. Moreover, an absorption \texttt{edge} was required to model features around 1.8 keV, potentially instrumental origin. The edge energy ($E_{\rm Edge}$) and optical depth ($\tau$) were free parameters.

This model provided a significantly improved description of the data, yielding $\chi^2/\mathrm{dof} = 688.3/642 \approx 1.07$ for Epoch 1 and $\chi^2/\mathrm{dof} = 646.7/644 \approx 1.00$ for Epoch 2. Given the high statistics of the data, these fits are considered acceptable. The best-fit models overlaid on the unfolded spectra are shown in Fig.~\ref{fig:broadband}.

For each broadband spectrum, we applied the Goodman–Weare Markov chain Monte Carlo (MCMC) algorithm implemented in Xspec to investigate the uncertainties of the best-fit parameters. We run MCMC simulations applying 200 walkers, a chain length of $10^7$, and a burn-in length of $10^5$.  The best-fit parameters and their $1\sigma$ confidence intervals derived from the MCMC posterior distributions are presented in Table~\ref{table:spec}.

The persistent emission is clearly detected up to 80 keV. The derived unabsorbed bolometric fluxes (1--250 keV) are very similar between the two epochs: $(3.85\pm0.01)\times 10^{-9}\,\rm{erg~s^{-1}~cm^{-2}}$ for Epoch 1 and $(3.81\pm0.01)\times 10^{-9}\,\rm{erg~s^{-1}~cm^{-2}}$ for Epoch 2, indicating slightly decreased emission during this period. Note that these estimated flux are $\sim30\%$ higher than the values solely from \nicer\ spectra in Sect.~\ref{sec:nicer_spec}.  The Galactic absorption column density ($N_{\rm H} \approx 2.6 \times 10^{22}\,\rm{cm^{-2}}$) is consistent between epochs and broadly agrees with the results from \nicer-only fits (Sect.~\ref{sec:nicer_spec}).

Most parameters of the continuum components (\texttt{nthcomp} and \texttt{relxillCp}) are consistent within $1\sigma$ uncertainties between these two epochs. Key reflection parameters are well-constrained, yielding $\Gamma \sim 1.9$, $\log(\xi / \rm{erg~cm~s^{-1}}) \sim 3.7$, $\log(n_{\rm e} / \rm{cm^{-3}}) \sim 16$, $A_{\rm Fe} \sim 1$, and a reflection fraction $f_{\rm refl.} \sim -(0.7 \text{--} 0.9)$. However, the coronal electron temperature ($kT_{\rm e}$ in \texttt{relxillCp}), the disk inclination ($i$), and the inner disk radius ($R_{\rm in}$), are poorly constrained, especially in Epoch 1 which has shorter exposures of \nustar\ spectra. The inclination derived for Epoch 2 ($i = 68^\circ \pm 5^\circ$) is consistent with the value of $\approx 74^\circ$ inferred from X-ray polarimetry \citep{Papitto25}. Fixing the inclination to $74^\circ$ in the Epoch 1 fit resulted in only a minor increase in $\chi^2$. The best-fit coronal temperatures, $kT_{\rm e}$, were  $225 \pm 115$ keV for Epoch 1 and $53 \pm 10$ keV for Epoch 2. The notably higher central value and significantly larger uncertainty for $kT_{\rm e}$ in Epoch 1 are primarily due to the shorter \nustar\ exposure available for this epoch.  The \texttt{constant} are close to unity for Epoch 1. For Epoch 2, the \texttt{constant} for \nustar/FPMA and FPMB relative to \nicer\ and \hxmt\ were 0.74 and 0.76, respectively. These values, indicating a $\sim25\%$ lower relative flux for \nustar, are attributed to the partial overlap between the observations; the \nicer\ and \hxmt\ exposures covered only the initial, brighter phase of the longer \nustar\ observation during which the source flux was decaying.

Our broadband spectral analysis can be compared with the recent results from \citet{Malacaria25}, who analyzed the joint \xmm\ and \nustar\ observations. Both studies confirm that the source showed strong relativistic reflection from the inner accretion disk. However, \citet{Malacaria25} adopted different continuum and reflection models. These fundamental methodological differences likely drive the divergence in several key physical parameters. While both studies confirm the presence of strong relativistic reflection, we found a highly ionized disk ($\log(\xi / \rm{erg~cm~s^{-1}}) \sim 3.7$) and an inclination of $i = 68^\circ \pm 5^\circ$. These differ with their report of a moderately ionized disk ($\log \xi \simeq 2.3$) at a lower inclination of $\sim 53^\circ$. Furthermore, they detect a $\sim 9.7$~keV absorption feature, interpreted as an ultra-fast outflow, which was not included in our model. These comparisons highlight that the choice of instrumentation and reflection model can significantly influence the derived physical parameters.

\section{Discussion and conclusion\label{sec:diss}}

In this work, we performed broadband timing and spectral analyses of the newly discovered AMXP \psrtar. X-ray pulsations were significantly detected across the $\sim$ 1.5--90 keV energy band as observed by \nicer, \ixpe, \hxmt\ and \Integ. 

\subsection{The magnetic field of \psrtar}\label{sec:mag}

X-ray pulsations from \psrtar\ were detected from the outburst peak down to MJD~60384.  During this period, the bolometric flux measured by \hxmt\ and \nicer\ varied over the range  $(0.1-4.2)\times10^{-9}~{\rm erg~s^{-1}~cm^{-2}}$. Adopting the distance of 10 kpc to the source \citep{Fu25}, the bolometric fluxes correspond to the mass accretion rate of $(0.14-5.58)\times 10^{17}~{\rm g~s^{-1}}$, by using the relation $L_X=4\pi D^2 F=\eta \dot{M}c^2$ where the accretion efficiency $\eta$ is set 0.1 for NS \citep{frank02}. The mass accretion rate is converted to $0.01-0.44\dot{M}_{\rm Edd}$, where $\dot{M}_{\rm Edd}=2\times10^{-8}~M_{\odot}~{\rm yr^{-1}}$ is the Eddington critical accretion rate.   The continuous detection of pulsations, even near the outburst peak ($\dot{M}_{\rm max} \approx 0.44\,\dot{M}_{\rm Edd}$), requires that the NS magnetic field is strong enough to truncate the accretion disk above the stellar surface. Using the relation $B = \mu/2R_{\rm NS}^3$, it sets a lower limit on the NS magnetic field \citep{Psaltis99b},
\begin{equation}\label{equ:mu_min}
\begin{split}
     B_{\rm min}=3.8\times10^{7}\gamma_{B}^{-1/2}\left(\frac{M_{\rm NS}}{1.4M_\odot}\right)^{1/4}\\
    \times \left(\frac{\dot{M}_{\rm max}}{0.44\dot{M}_{\rm Edd}}\right)^{1/2}\left(\frac{R_{\rm NS}}{11~{\rm km}}\right)^{-5/4}~{\rm G},
\end{split}
\end{equation}
where $M_{\rm NS}$ and $R_{\rm NS}$ are the NS mass and radius, respectively. The parameter $\gamma_B$ is defined as $\gamma_B \equiv (B_\phi/B_p)(\Delta r_0/r_0)$, where $B_p$ and $B_\phi$ are the poloidal and toroidal components of the magnetic field, $\Delta r_0$ is the radial width of the interaction region, and $r_0$ is the disruption radius of the disk flow \citep{Ghosh78}.  This factor is not well constrained and is assumed to be in the range 0.01--1 \citep{Psaltis99b}. If the values of $M_{\rm NS}=1.4M_{\odot}$, $R_{\rm NS}=11$ km, $\gamma_B=1$, and $\dot{M}_{\rm max}=4.58\times 10^{17}~{\rm g~s^{-1}}$ are substituted to Equation~(\ref{equ:mu_min}), the minimum magnetic field is $3.8\times10^{7}~{\rm G}$. At the lowest accretion rate with the pulsation detected, it corresponds to the upper limit of the  polar magnetic field via the relation,
\begin{multline}
\label{equ:mu_max}
     B_{\rm max}=1.6\times10^{9}\left(\frac{\gamma_{B}}{0.01}\right)^{-1/2}\left(\frac{\dot{M}_{\rm min}}{0.01\dot{M}_{\rm Edd}}\right)^{1/2}\\
     \times\left(\frac{M_{\rm NS}}{2.3M_\odot}\right)^{5/6}\left(\frac{R_{\rm NS}}{11~{\rm km}}\right)^{-5/2}\left(\frac{\nu}{447.87~{\rm Hz}}\right)^{-7/6}~{\rm G},
\end{multline}
where, $\nu$ is the AMXP spin frequency, $\dot{M}_{\rm min}$ is the minimum accretion rate when the pulsation has been detected. We take $\gamma_B=0.01$ and $M_{\rm NS}=2.3M_{\odot}$ \citep[see e.g.,][for the possibility of massive NS]{Freire08} to obtain the upper limit of the magnetic field, $B_{\rm max}=1.6\times10^{9}~{\rm G}$. 

A more direct estimate of the magnetic field can be obtained from the observed spin-up rate. During the interval MJD 60361--60377, \psrtar\ exhibited a significant spin-up $\dot{\nu}$ of $(3.15 \pm 0.36) \times10^{-13}~{\rm Hz~s^{-1}}$. We estimated the average bolometric flux during this specific epoch by interpolating the available flux from \nicer, yielding the average value of $\sim2.67 \times 10^{-9}~{\rm erg~cm^{-2}~s^{-1}}$.  Due to the sparse observational coverage and lacking of hard X-ray band coverage, the uncertainty on this average flux is difficult to estimate reliably; we therefore adopt a  uncertainty estimate of 30\%. This average flux corresponds to a mass accretion rate $\dot{M}$ of $(0.28\pm0.03)\dot{M}_{\rm Edd}$. Assuming the spin-up is solely due to the accretion torque transferring angular momentum from the disk to the NS, the magnetic field strength can be estimated using the relation \citep{Shapiro83, Tong15, Pan22}, 
\begin{equation}\label{equ:spinup}
\begin{split}
B=0.21\left( \frac{I}{1.5\times 10^{45}~{\rm g~cm^2}}\right)^{7/2}\left( \frac{R_{\rm NS}}{11~{\rm km}}\right)^{-3}\left( \frac{M_{\rm NS}}{1.4M_\odot}\right)^{-3/2}\\
\times\left( \frac{\dot{\nu}}{3.15\times10^{-13}~{\rm Hz~s^{-1}}}\right)^{7/2}\left( \frac{\dot{M}}{0.28\dot{M}_{\rm Edd}}\right) ^{-3}\times10^8~{\rm G},
\end{split}
\end{equation}
where $I=1.5\times 10^{45}~{\rm g~cm^2}$ is the NS moment of inertia  \citep[see e.g.,][]{Worley08}. Accounting for the uncertainties in $\dot{\nu}$ and $\dot{M}$, we obtain a magnetic field strength of $(2.1 \pm 2.0) \times 10^7$~G.  Combined with the lower limit inferred from Equation (\ref{equ:mu_min}), the magnetic field strength is $\sim4\times10^7$ G.

\subsection{The burst induced pulse profile variation}

During its 2024 outburst, \psrtar\ exhibited frequent type I X-ray bursts with a recurrence time increasing from approximately 1.55 to 10 hr as the persistent emission decreased.  The extensive burst dataset collected by \hxmt\ (60 bursts), \nustar\ (23 bursts), and \ixpe\ (52 bursts) provides an excellent opportunity to investigate variations in the pulse profile shape during these events across a broad energy range (2--60\,keV). 

A previous analysis by \citet{Molkov24} reported significant differences between the burst and persistent pulse profiles. However, that study utilized an initial orbital and spin ephemeris derived from early outburst observations \citep{2024ATel16464}. Applying the refined timing solution presented in this work (Table~\ref{table:eph}) is crucial for accurately folding the burst data and characterizing any intrinsic pulse profile changes. We generated stacked, phase-folded profiles for pre-burst, post-burst, and burst epochs using data from \ixpe\ (2--8\,keV), \nustar\ (3--10, 10--35, 35--60\,keV), and \hxmt/ME/HE (5--30 and 20--60\,keV), as detailed in Section~\ref{sec:tmbur}.

Our main findings are as follows.

\begin{itemize}
    \item The pulsations during bursts from \psrtar\ are significantly detected by \hxmt/ME/HE, \ixpe, and \nustar in 2--60 keV.
    \item The first overtone component of the pulse profiles was shown in \hxmt/HE and more evident in ME. 
    \item The ratio of the amplitudes of fundamental ($A_1$) to the unpulsed amplitude ($A_0$) during the bursts, are smaller than the pre- and post-bursts in \ixpe. However, the ratio $A_1/A_0$ during the bursts are higher than the pre- and post-bursts in \hxmt/ME/HE.
    \item The pulse profiles for pre-, post-, and during bursts are similar. However, the pulse profiles during X-ray bursts lag behind the pre- and post-burst pulse profiles, $\Delta \phi\approx0.15$, 0.11, and 0.02 for \ixpe, \hxmt\ ME, and HE, respectively, and $\Delta \phi\approx0.21$, 0.10, and 0.07 for \nustar\ in 3--10, 20--35, and 35--60 keV, respectively. It suggests a smaller time lag at higher energy band.
\end{itemize}

The broadband X-ray pulsation of AMXPs is explained in two component emission, the thermal emission (below 8 keV) from the hot spot(s) on NS surface and the Componization component from the accretion column. When a type I X-ray burst occurs, it produces nearly isotropic thermal emission on the NS surface with the photon energy below 20 keV. The detected pulsation during bursts indicates that the X-ray burst emission did not destroy or screen the hot spot and accretion column.  It is more likely that the bursts produce hotter thermal emission on the NS surface, resulting in a smaller ratio between pulsed and unpulsed emission in soft X-ray band, as observed in \ixpe. 

The phase lag observed in \psrtar\ is analogous to the similar phenomenon observed in GRO J1744--28 during its type II X-ray bursts. 
Such bursts in GRO J1744--28 were induced from the increased accretion rate, which is likely driven by disk instabilities. The accreted matter was channeled along a different set of field lines to fall onto NS surface around the polar cap. Therefore, the location of the hot spot was shifted, resulting in a phase lag as proposed by \citet{Miller96}. 

We suggest that a related mechanism can also explain the phase lag during type I X-ray bursts in \psrtar. The intense burst radiation can significantly interact with the inner accretion disk via the Poynting-Robertson drag, potentially enhancing the mass accretion rate onto the NS surface temporarily. This radiation-induced enhancement of accretion, analogous to the intrinsic accretion increase in GRO J1744--28's type II X-ray bursts, could similarly alter the geometry or location of the accreting material impacting the magnetic footprint associated with the persistent emission. This shift in the effective emission center would manifest as the observed phase lag relative to the pre- and post-burst profiles.

The energy dependence of the phase lag can potentially be understood within the framework of a stratified emission region. If the lower-energy X-rays (including the thermal hot spot and the added burst emission) originate closer to the NS surface where the footprint shift is most pronounced, while the higher-energy emission (likely dominated by Comptonization in the accretion column above the surface) originates from regions less affected by the precise surface impact geometry, then the phase lag would naturally decrease with increasing energy, as observed.

\begin{acknowledgments}
We thank the referee for the constructive and valuable comments which improve the manuscript. This paper is dedicated to the memory of Prof. Dr. Maurizio Falanga, whose contributions to X-ray astrophysics continue to inspire us.
This work was supported the Major Science and Technology Program of Xinjiang Uygur Autonomous Region (No. 2022A03013-3), and the Science and Technology program of
Hunan Province (No. 2024JC0001). Z.S.L. and Y.Y.P. were supported by National Natural Science Foundation of China (12273030, 12103042).  This work made use of data from the \hxmt\
mission, a project funded by China National Space Administration (CNSA) and the Chinese Academy of Sciences (CAS), and also from the High Energy Astrophysics Science Archive Research Center (HEASARC), provided by NASA’s Goddard Space Flight Center. EP
is a space mission supported by Strategic Priority Program on Space Science
of Chinese Academy of Sciences, in collaboration with ESA, MPE and
CNES.
\end{acknowledgments}

\bibliography{pulsars}{}
\bibliographystyle{aasjournal}

\end{document}